\documentstyle{mn}

\title[Radio properties of radio-quiet quasars]{The radio properties of radio-quiet quasars}

\author[M.J. Kukula et al.]
{Marek~J.~Kukula$^{1,2}$, James~S.~Dunlop$^2$, David~H.~Hughes$^2$ and Steve~Rawlings$^3$\\
$^1$~Space Telescope Science Institute, 3700 San Martin Drive, Baltimore MD 21218, U.S.A.\\
$^2$~Institute for Astronomy, Department of Physics and Astronomy,
University of Edinburgh, Blackford Hill, Edinburgh EH9 3HJ, U.K. \\
$^3$~Department of Astrophysics, Nuclear \& Astrophysics Laboratory,
University of Oxford, Keble Road, Oxford OX1 3RH, U.K. \\}

\date{}

\begin{document}
\maketitle

\begin{abstract}

Although radio-quiet quasars (RQQs) constitute
$\stackrel{>}{\sim}90\%$ of optically-identified quasar samples their
radio properties are only poorly understood. In this paper we present
the results of a multi-frequency VLA study of 27 low-redshift RQQs.
We detect radio emission from 20 objects, half of which are unresolved
($\leq 0.24''$). In cases where significant structure can be resolved,
double, triple and linear radio sources on scales of a few kpc are
found.  The radio emission (typically) has a steep spectrum ($\alpha
\sim 0.7$, where $S \propto \nu^{-\alpha}$), and high brightness
temperatures ($T_{B} \geq 10^{5}$~K) are measured in some of the radio
components. The RQQs form a natural extension to the radio luminosity
- absolute magnitude distribution of nearby Seyfert 1s. We conclude
that a significant fraction of the radio emission in RQQs originates
in a compact nuclear source directly associated with the quasar. There
are no significant differences between the radio properties of RQQs
with elliptical hosts and those in disc galaxies within the current
sample.
\end{abstract}


\section{Introduction}

Shortly after their initial identification in 1963 by Schmidt it
became clear that less than 10\% of quasars are strong sources of
radio emission.  The gap in the quasar radio luminosity function is
extremely pronounced, with only a handful of objects occupying the
region between quasars which are radio-loud and those which are
radio-quiet (Kellermann et al. 1989; Miller, Peacock \& Mead 1990;
Miller, Rawlings \& Saunders 1993, henceforth MRS93), and it is now
widely accepted that there are two distinct populations of quasar.

However, from $\sim100~\mu$m through to soft X-ray wavelengths the
properties of radio-loud and radio-quiet quasars (RLQs and RQQs
respectively) are generally very similar, to the extent that to first
order an RLQ spectrum can be considered to be the spectrum of an RQQ
with the addition of a strong power-law component in the radio. The
presence or absence of this radio component must be a fundamental
indicator of the processes occurring in the quasar, but there is still
no consensus as to exactly what it signifies.

Studies of less luminous, nearby active galaxies have shown that the
radio-loud objects ({\it ie} Radio Galaxies) are invariably elliptical
systems whereas the radio-quiets (Seyferts) tend to be spirals,
suggesting that some property of gas-rich disc galaxies inhibits the
formation of large, powerful radio sources (see Osterbrock 1991).  Two
factors have encouraged the extension of this result to the higher
redshifts and larger nuclear luminosities typical of quasars: the
success of Unified Schemes which link Radio Galaxies and RLQs via
beaming effects and viewing angle (see Urry \& Padovani 1995) and the
established fact that Seyfert 1 nuclei and RQQs form a continuous
sequence in terms of their optical luminosities and have identical
emission line characteristics (see Osterbrock 1991, Antonucci 1993).

However, recent advances in ground- and space-based observing have
allowed the host galaxies of nearby quasars to be imaged and {\it
reliably} classified for the first time. The results have shown that
{\it not all} RQQs lie in disc systems but that as many as 50\% might
be found in elliptical hosts ({\it eg} Taylor et al. 1996, Disney et
al. 1995, Lacy, Rawlings \& Hill 1992, V\'{e}ron-Cetty \& Woltjer
1990). Indeed, there is some evidence that elliptical galaxies might
account for all of the most optically luminous RQQs (Taylor et
al. 1996). Clearly the simple `radio-loud $\equiv$ elliptical,
radio-quiet $\equiv$ disc' picture can no longer be supported, and it
has become more important than ever to determine in what respects
radio-quiet quasars are different from their radio-loud
counterparts. Unfortunately, the most obvious wavelength regime in
which they differ - the radio - is also the regime in which least is
known about the properties of RQQs. Most radio surveys of
optically-selected quasar samples have lacked the sensitivity to
detect the radio-quiet objects, and the few high-sensitivity surveys
have tended to give only fluxes at a single frequency, with only
limited information on radio structure and spectral index.

Like Seyferts, RQQs are not radio silent and with sensitivities at the
milliJansky level modern multi-element synthesis telescopes can easily
detect the radio emission from many of the nearer ($z \leq 0.3$)
objects.  In this paper we present the results of a multi-frequency,
high-resolution radio survey of low-redshift ($0.03 \leq z \leq 0.3$),
low-luminosity (M$_{V}>-26$) RQQs using the Very Large Array (VLA) in
its high-resolution A-configuration.  The aim of this work was to
search for compact, non-thermal radio sources associated with the
quasars, to determine their properties and to compare them with
those of radio-loud objects.

As well as our own observations, we also make use of data from the
literature.  Our RQQ sample comprises 27 objects, 19 of which have
already been observed with A-array at 4.8~GHz by Kellermann et
al. (1989) as part of their radio study of the Bright Quasar Survey
(BQS; Schmidt \& Green 1983, Green, Schmidt \& Liebert 1986). We have
observed all 27 RQQs at 1.4~GHz and completed the coverage of the
sample at 4.8~GHz with observations of the 8 non-BQS quasars. In
addition we obtained 8.4-GHz maps for 21 objects.

Seventeen of the RQQs in our sample were also included in a major
investigation into quasar host galaxies (Dunlop et al. 1993, Taylor et
al. 1996). We use these objects to investigate the relationship
between the host and the `radio loudness' of the quasar and refer
to them as our `host galaxy subsample'. 

Two objects (PG~0007$+$106 \& 1635$+$119) have unusually large radio
luminosities and arguably are not true {\it radio-quiet} quasars but
belong to a separate class of their own. We use the term
`radio-intermediate quasar' (RIQ) to describe these.

The paper is structured as follows. Data acquisition and reduction are
discussed in Section 2, followed by a description of the resulting
radio images in Section 3. Section 4 of the paper summarises the
debate over the origin of non-thermal radio emission in RQQs and
Section 5 explores the implications of the current observations, and
compares the RQQ emission with the radio properties of other types of
active galactic nuclei (AGN).  In Section 6 we examine the
relationship between the radio properties of the 17 RQQs in our host
galaxy subsample and the properties of the hosts themselves, and our
conclusions are summarised in Section 7. We assume $H_{0}
=50$~kms$^{-1}$Mpc$^{-1}$ and $\Omega_{0} = 1$ throughout.

\section{Observations and data reduction}

In order to build up a picture of RQQ radio emission over a range of
frequencies, snapshot observations were made with the VLA in
A-configuration at 1.4, 4.8 and 8.4~GHz ($L$, $C$ and $X$ bands
respectively).

Many of the objects in our sample have already been observed at one
or more of these frequencies as part of previous surveys and we make
use of these data in the current study.

All of the Palomar-Green (PG) objects were observed at 4.8~GHz with
the VLA in both D- and A-configurations as part of the BQS (November
1983, Kellermann {\it et al.} 1989; June 1991, MRS93).  PG~0050+124
(I~Zw~1) was observed in A-configuration at 8.4~GHz (July 1991, Kukula
et al. 1995) as part of the CfA Seyfert sample.

We also include additional data at 1.4~GHz for three quasars
(PG~0007+106, PG~0026+129 \& PG~0052+251). These maps have been
presented previously in Philip Miller's PhD thesis (Miller 1992), and
we publish them here with his permission. The observations were made
in August 1991.

Our own 1.4- and 8.4-GHz observations were made with A-configuration
in December 1992 and July 1995. 4.8-GHz observations were made of the
eight non-PG quasars in July 1995. Table~1 lists the observing dates
for all 27 objects at each frequency to allow an assessment of the
potential impact of any source variability on the individual flux
measurements.

Data reduction followed the usual procedure within {\sc aips}. Two
50-MHz IF channels were used and the flux densities were calibrated
relative to 3C48, to give an estimated flux uncertainty of 5\% in the
final maps.  A natural weighting scheme was used for the Fourier
transform, giving noise levels in the final maps of $\sigma \sim$ 300,
80 and 60 $\mu$Jy~beam$^{-1}$ at 1.4, 4.8 and 8.4~GHz respectively and
synthesized beamwidths of 1.4, 0.4 and 0.24$''$ (FWHM).
Self-calibration was used on the brightest sources. We estimate the
positional information in the high-resolution 8.4-GHz maps to be
accurate to within $\sim 50$~mas.

\begin{table}\label{dates}
\caption{Dates of VLA A-array observations (month/year). Entries
marked with an asterisk ($^{*}$) refer to the observations by
P. Miller. Additional 1.4~GHz observations of these sources were made
in 12/92.}

\centering
\begin{tabular}{rlll}
\hline
\multicolumn{1}{c}{Object} & 1.4~GHz & 4.8~GHz & 8.4~GHz \\ \hline
PG 0007+106 & 08/91$^{*}$ & 11/83 & 07/95 \\
PG 0026+129 & 08/91$^{*}$ & 11/83 &  -- \\
0046+112    & 12/92 & 07/95 & 07/95 \\
PG 0050+124 & 07/95 & 07/95 & 07/91 \\
PG 0052+251 & 08/91$^{*}$ & 11/83 & 12/92 \\
0054+144    & 12/92 & 07/95 & 07/95 \\
PG 0157+001 & 12/92 & 11/83 & 12/92 \\
0244+194    & 12/92 & 07/95 & 07/95 \\
0257+024    & 12/92 & 07/95 & 07/95 \\
PG 0804+761 & 12/92 & 11/83 & 12/92 \\
PG 0921+525 & 12/92 & 11/83 & 12/92 \\
PG 0923+201 & 07/95 & 11/83 & 07/95 \\
PG 0953+414 & 12/92 & 06/91 & 07/95 \\
PG 1012+008 & 12/92 & 11/83 & 12/92 \\
PG 1116+215 & 12/92 & 11/83 & 12/92 \\
PG 1149$-$110&12/92 & 11/83 & -- \\
PG 1211+143 & 12/92 & 11/83 & -- \\
PG 1402+261 & 12/92 & 11/83 & -- \\
PG 1440+356 & 12/92 & 11/83 & -- \\
1549+203    & 12/92 & 07/95 & 07/95 \\
PG 1612+261 & 12/92 & 11/83 & 12/92 \\
PG 1613+658 & 12/92 & 11/83 & 12/92 \\
1635+119    & 12/92 & 07/95 & 07/95 \\
PG 1700+518 & 12/92 & 11/83 & 12/92 \\
PG 2130+099 & 12/92 & 11/83 & 12/92 \\
2215$-$037  & 12/92 & 07/95 & 07/95 \\
2344+184    & 12/92 & 07/95 & 07/95 \\
\hline
\end{tabular}
\end{table}

\begin{table*}\label{optical}
\caption{Redshifts, $V$ magnitudes and optical positions of the RQQs
in the current sample. Optical positions for the PG objects are taken
from MRS93; the remaining positions, and
values for $z$ and $V$, are from V\'{e}ron-Cetty \& V\'{e}ron
(1993). All positions are given in B1950 co-ordinates, but the
accuracy varies from object to object.}

\centering
\begin{tabular}{rccccl}
\hline
Object      & $z$ & $V$ & \multicolumn{2}{c}{Optical position (B1950)}& Alternative names \\ 
            & & & RA {\it h m s} & Dec $^{\circ}~'~''$    &             \\ \hline   
PG 0007+106 & 0.089 & 15.40 & 00 07 56.74 &+10 41 48.3    & III Zw 2    \\          
PG 0026+129 & 0.142 & 15.41 & 00 26 38.07 &+12 59 29.6    &             \\          
0046+112    & 0.280 & 17.10 & 00 46 55.50 &+11 12 06.1    & PHL850      \\          
PG 0050+124 & 0.061 & 14.03 & 00 50 57.8~ &+12 25 20.0    & I Zw 1      \\          
PG 0052+251 & 0.154 & 15.90 & 00 52 11.10 &+25 09 23.8    &             \\          
0054+144    & 0.171 & 15.71 & 00 54 31.95 &+14 29 58.4    & PHL909      \\          
PG 0157+001 & 0.164 & 15.69 & 01 57 16.30 &+00 09 09.5    & Mrk 1014    \\          
0244+194    & 0.176 & 16.66 & 02 44 51.70 &+19 28 23.8    & MS 02448+19 \\          
0257+024    & 0.115 & 16.10 & 02 57 53.91 &+02 29 00.8    & US3498      \\          
PG 0804+761 & 0.100 & 15.15 & 08 04 35.4~ &+76 11 32~~    &             \\          
PG 0921+525 & 0.035 & 16.0~ & 09 21 44.40 &+52 30 07.8    & Mrk 110     \\          
PG 0923+201 & 0.190 & 15.83 & 09 23 05.81 &+20 07 07.2    & Ton 1057    \\          
PG 0953+414 & 0.239 & 15.55 & 09 53 48.31 &+41 29 58.2    & K 438-7     \\          
PG 1012+008 & 0.185 & 15.85 & 10 12 20.78 &+00 48 33.0    &             \\          
PG 1116+215 & 0.177 & 15.04 & 11 16 30.1~ &+21 35 43~~    &             \\          
PG 1149$-$110& 0.049& 15.46 & 11 49 30.28 &$-$11 05 42.4  &           \\          
PG 1211+143 & 0.085 & 14.63 & 12 11 44.86 &+14 19 53.1    &             \\          
PG 1402+261 & 0.164 & 15.82 & 14 02 59.28 &+26 09 51.8    & Mrk 478     \\          
PG 1440+356 & 0.077 & 14.58 & 14 40 04.55 &+35 39 07.4    &             \\          
1549+203    & 0.250 & 16.50 & 15 49 49.41 &+20 22 56.6    & 1E15498+203, LB906, MS 15498+20\\
PG 1612+261 & 0.131 & 15.41 & 16 12 08.72 &+26 11 46.7    & Ton 0256    \\          
PG 1613+658 & 0.129 & 15.49 & 16 13 36.24 &+65 50 37.5    & Mrk 876     \\          
1635+119    & 0.146 & 16.50 & 16 35 25.90 &+11 55 46.4    & MC 2        \\          
PG 1700+518 & 0.290 & 15.43 & 17 00 13.36 &+51 53 36.2    &             \\          
PG 2130+099 & 0.063 & 14.64 & 21 30 01.18 &+09 55 01.1    & II Zw 136   \\          
2215$-$037  & 0.241 & 17.20 & 22 15 11.70 &$-$03 47 50.2  & MS 22152$-$03, EX2215$-$037	\\
2344+184    & 0.138 & 15.9~ & 23 44 53.30 &+18 28 18.0    & E2344+184   \\ \hline   

\end{tabular}
\end{table*}

\section{Results}

Table~2 lists the redshifts, $V$ magnitudes and optical positions for
the objects in the current sample.  Of the 27 RQQs 20 were detected at
1.4~GHz. Half of the detected RQQs are unresolved point sources at all
three frequencies, but ten objects (PG~0007+106, PG~0026+129,
PG~0052+251, PG~0157+001, PG~0804+761, PG~0921+525, PG~1612+261,
1635+119, PG~1700+518 and PG~2130+099) show evidence for extended
structure on arcsecond scales. Maps of eight of these sources are
shown in Figures 1 \& 2 with contour levels listed in
Table~3. PG~0804+761 is known to possess a double radio source at
4.8~GHz (MRS93) but the current maps at 1.4 and 8.4~GHz show only the
brightest of the two components. Gower \& Hutchings (1984) list
1635+119 as an extended double source, but we detect only the
unresolved core component with the current observations.

The radio positions of the detected RQQs, determined from 2D Gaussian
fits to the high-resolution 8.4-GHz images (or from the 1.4-GHz images
when the object was not observed in $X$-band), are given in Table~4
along with the total flux densities of each distinct radio component,
again determined from a 2-D Gaussian fit. Sources of error in these
flux values include calibration uncertainties of order $5\%$, but the
major contributor is usually the rms noise in the final map.  In the
case of non-detections, $3\sigma$ upper limits are given. Table~5
lists the derived radio luminosities, radio source sizes and
brightness temperatures.

Table~4 also lists spectral indices, $\alpha$ (where $S \propto
\nu^{-\alpha}$), for each radio component, calculated between 1.4, 4.8
and 8.4~GHz. The 4.8-GHz fluxes listed by MRS93 are {\it peak} values,
but these will only differ from the total flux when the radio
component is significantly resolved. 

Two further caveats apply here. The first is that the A-array beam
decreases in size with increasing frequency. The VLA therefore becomes
less sensitive to extended radio emission at higher frequencies (the
fact that such emission usually has a steep spectrum will exacerbate
this problem).  Consequently there is the possibility that faint,
diffuse emission which is included in the larger beam at low
frequencies might be resolved out at higher frequencies when the
smaller beam of the array is only sensitive to compact structures.
This would have the effect of making the radio spectrum appear steeper
than it actually is.  However, even at 8.4~GHz, A-array remains
sensitive to extended emission on angular scales of up to 7$''$.  This
is much larger than the synthesized beam at 1.4~GHz (1.4$''$)
and, in practice, we find that the 1.4-GHz emission from our RQQs is
unresolved in almost every case. It therefore seems reasonable to
conclude that the amount of diffuse emission picked up at 1.4~GHz but
resolved out at 8.4~GHz is unlikely to be significant ({\it ie}
`aperture corrections' are not required).

More seriously, observations at different frequencies were often
separated by a period of several years (see Table~1), during which the radio
brightness of the RQQ may have varied considerably. Hence the most
reliable spectral indices are those derived from measurements which
were made concurrently - in most cases this will be the value derived
between 1.4 and 8.4~GHz ($\alpha^{1.4}_{8.4}$ in Table~4).

\begin{figure*}
\vspace {16.0truecm}
 
\includegraphics{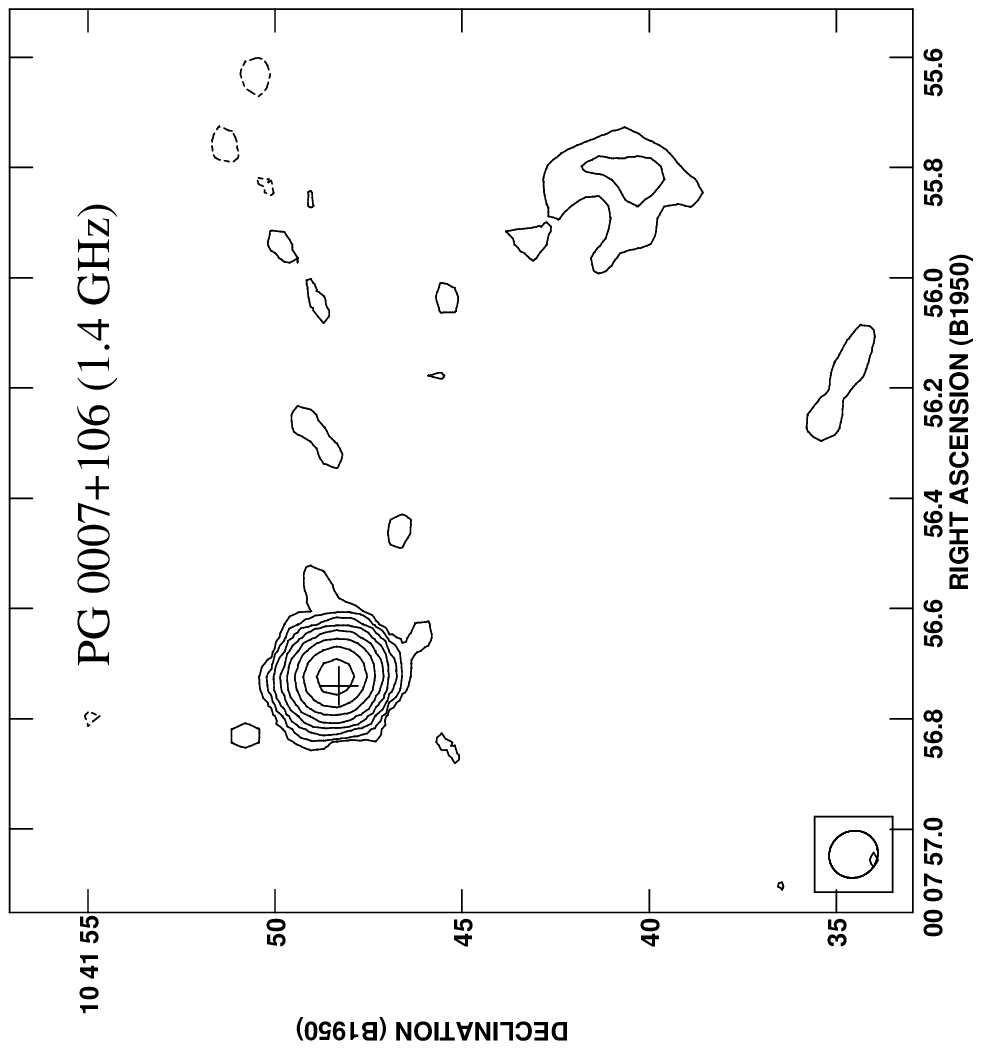}
\includegraphics{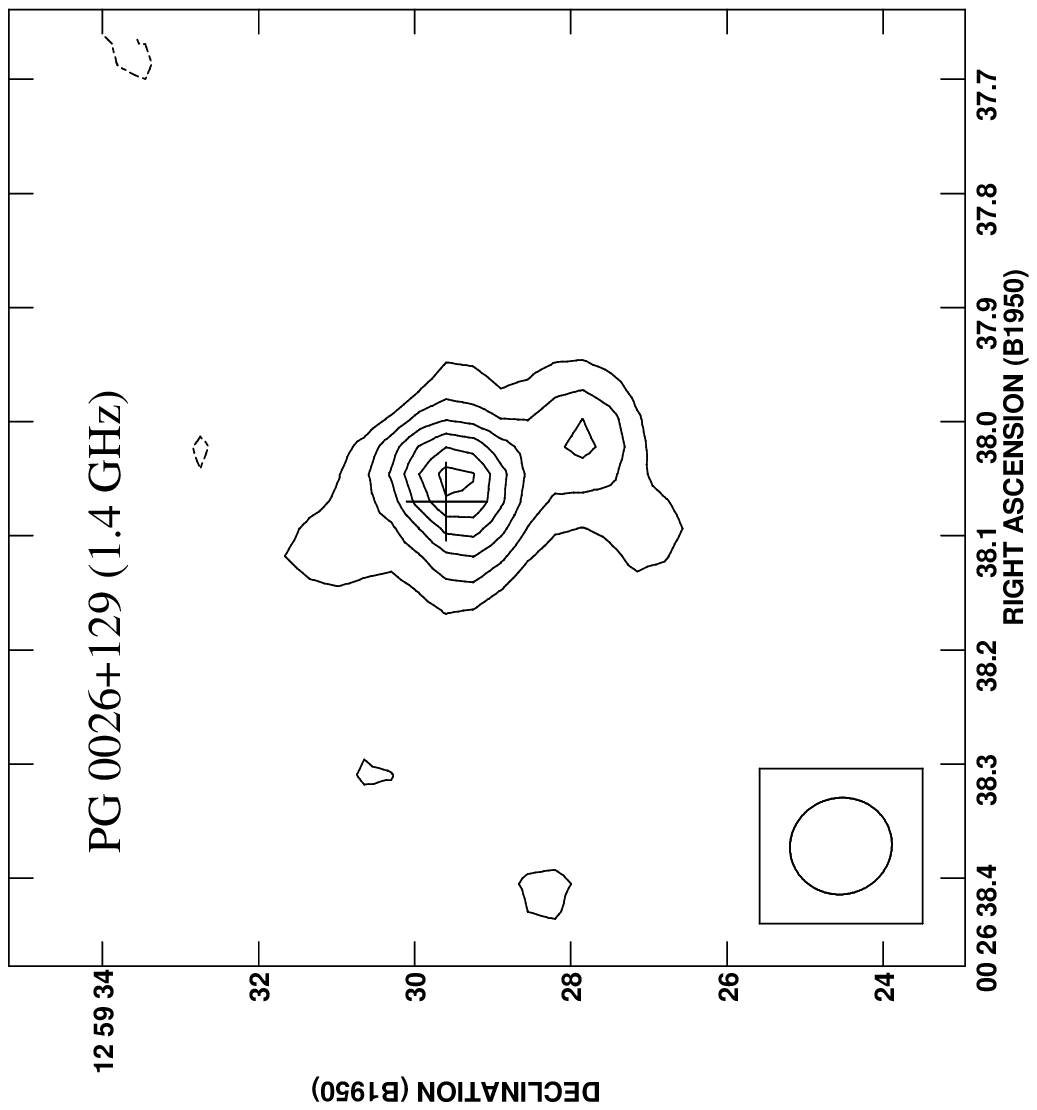}
\includegraphics{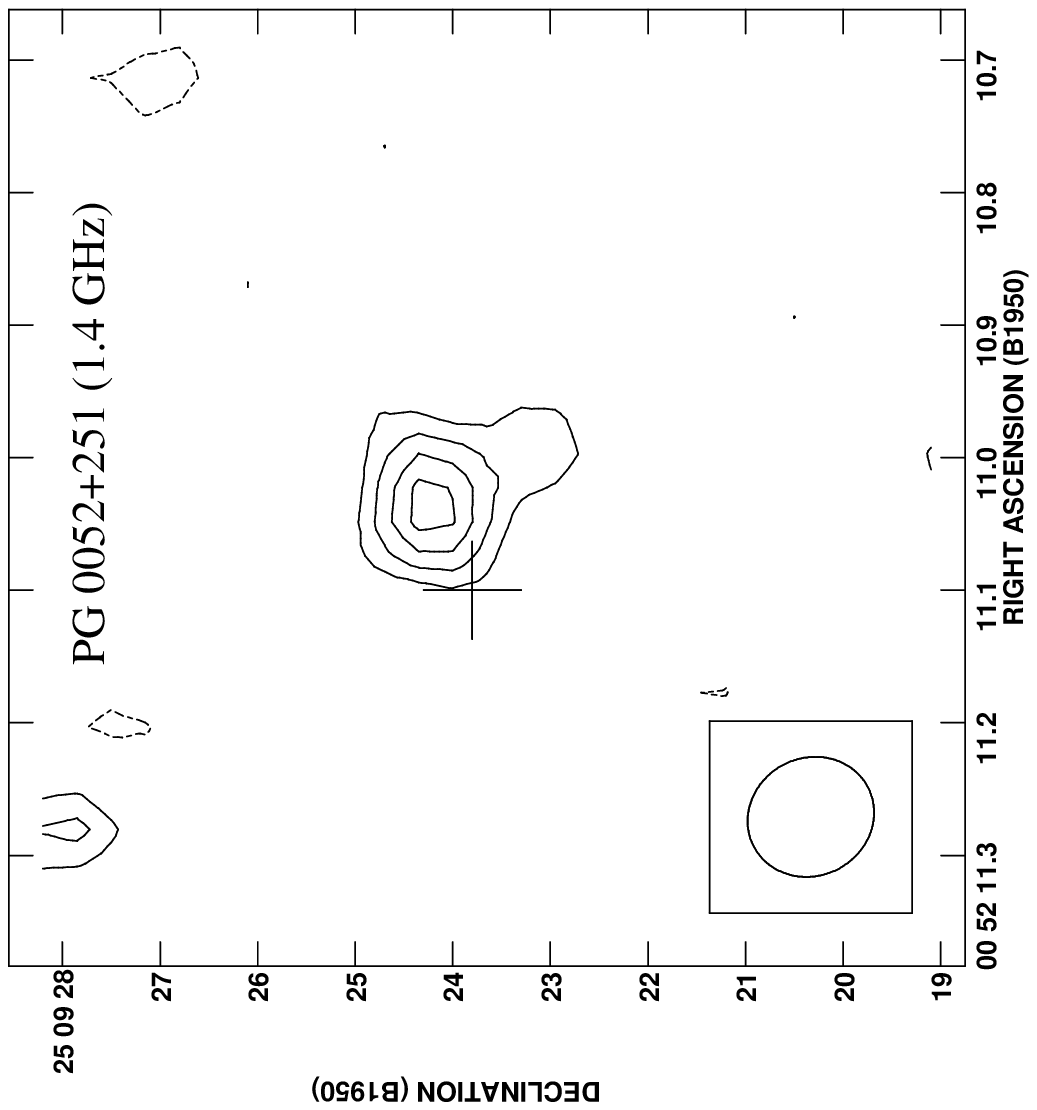}
\includegraphics{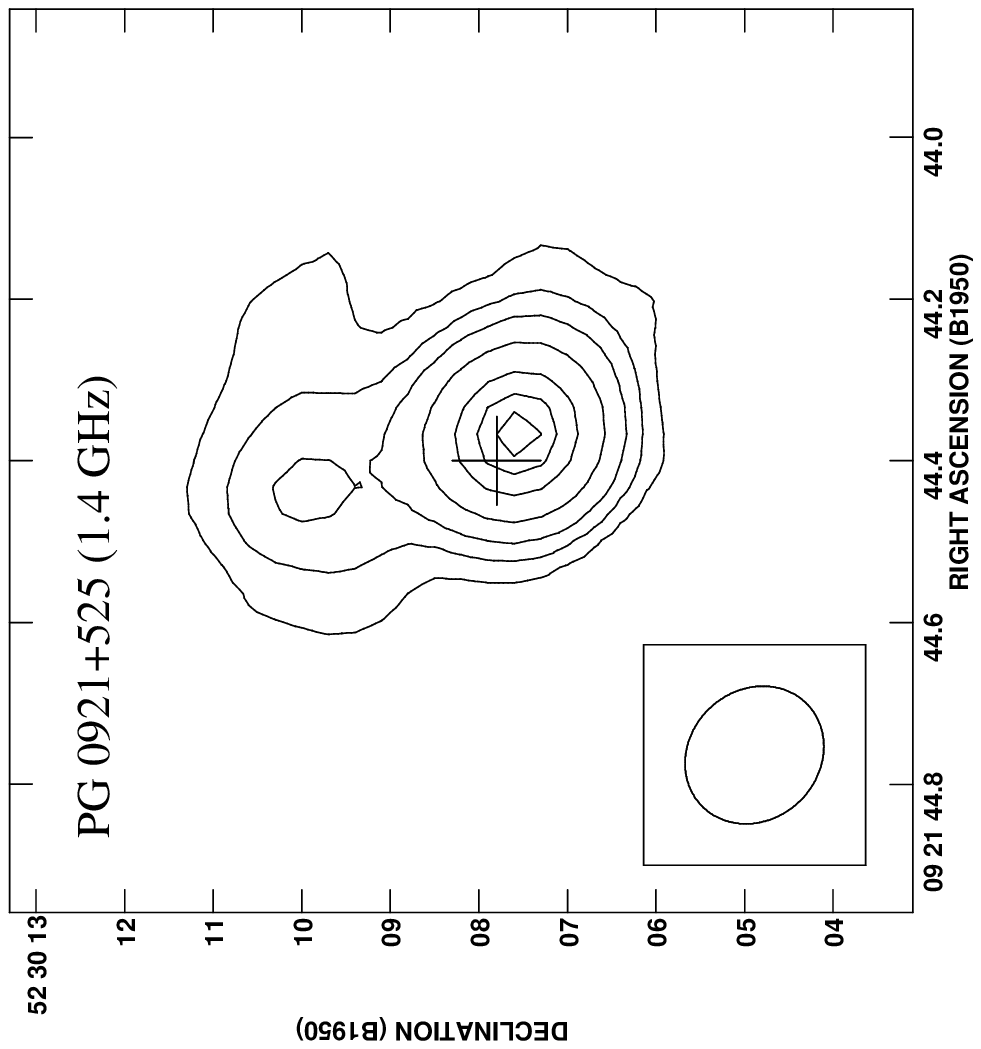}

\caption[]{Radio source structures at 1.4 GHz. From left to right:
PG~0007+106, PG~0026+129, PG~0052+251 \&
PG~0921+525. Crosses mark the position of the optical nucleus, scaled
to represent a nominal uncertainty of $\pm 0.5 ''$. The FWHM of the
{\sc clean} beam is indicated by the ellipse at the bottom left of
each frame. Contour levels are given in Table 3.}

\end{figure*}

\begin{figure*}
\vspace {15.5truecm}
 
\includegraphics{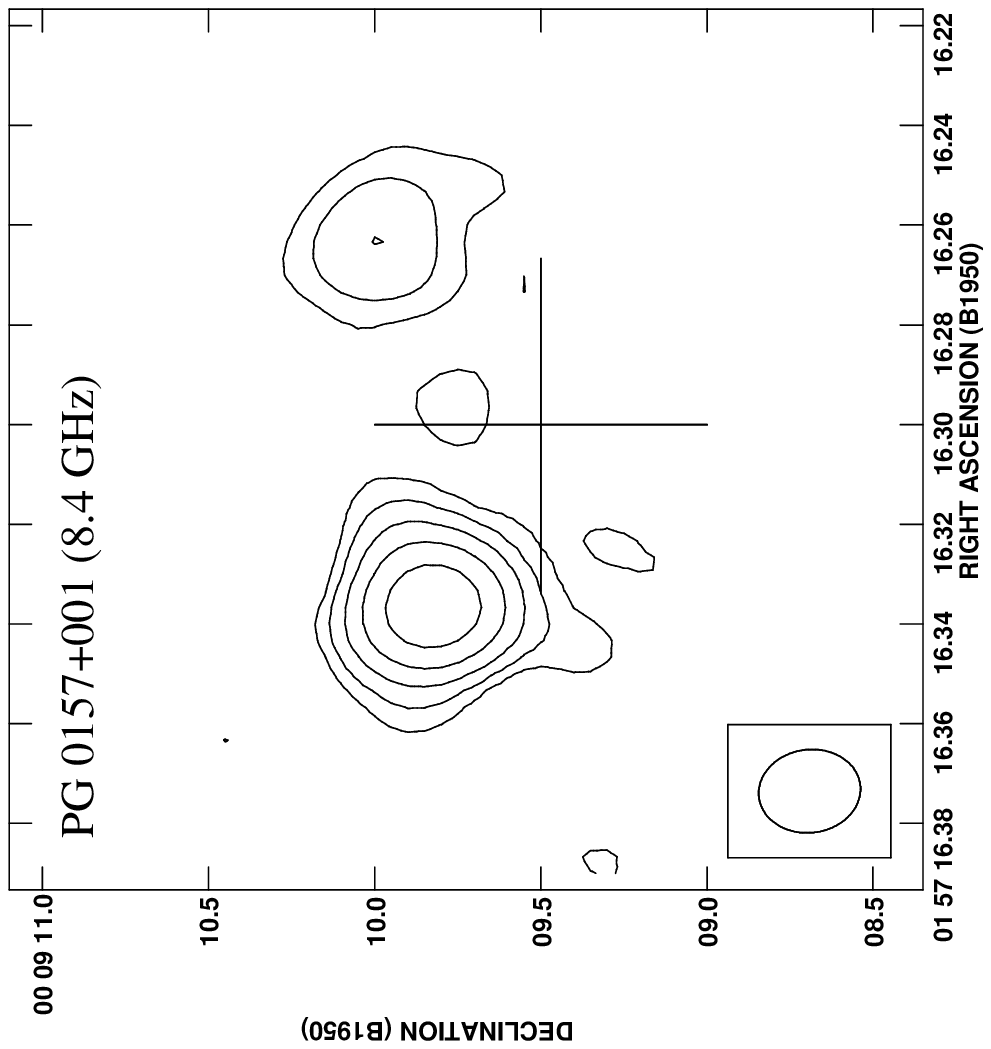}
\includegraphics{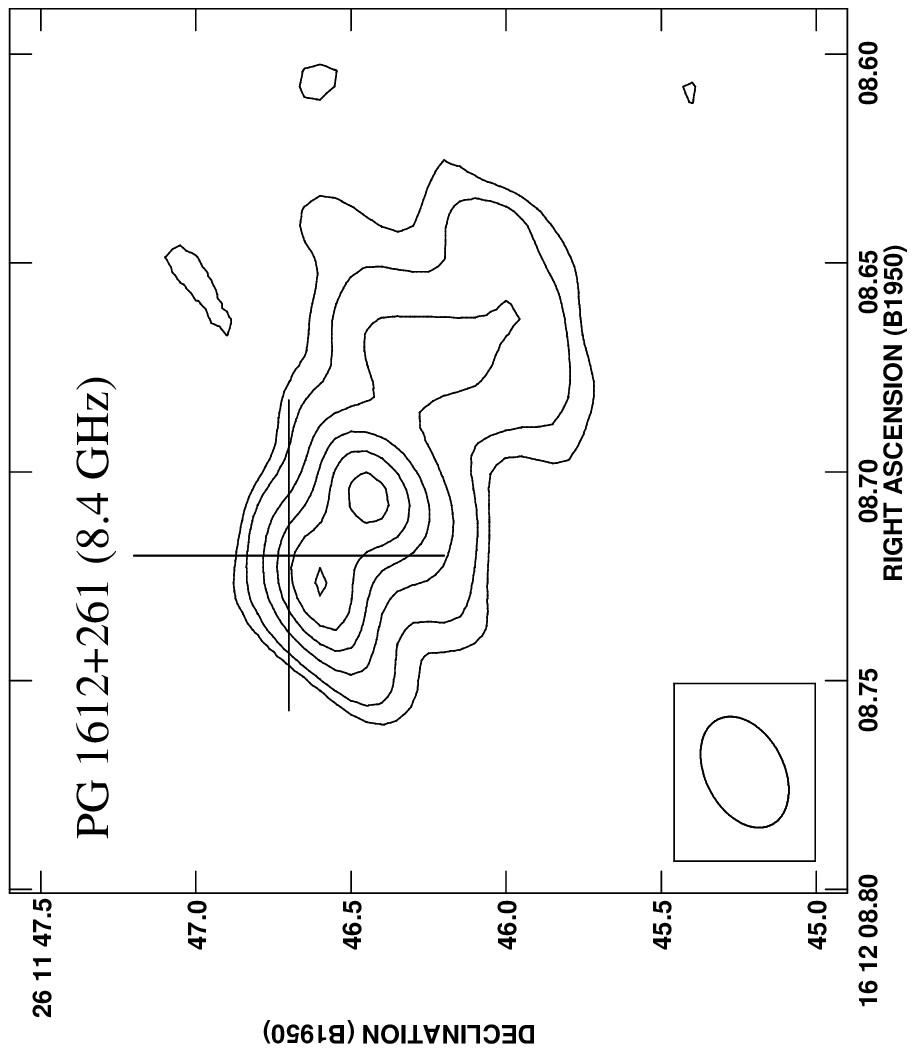}
\includegraphics{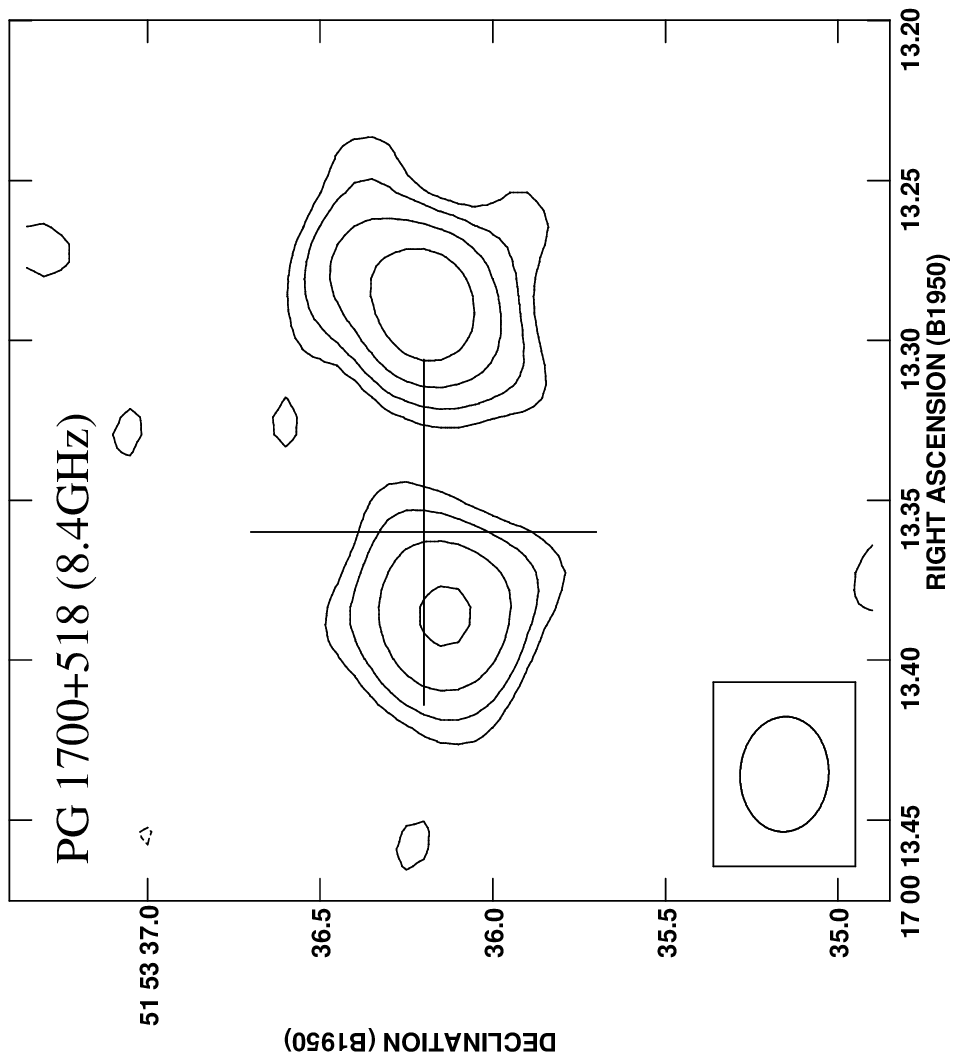}
\includegraphics{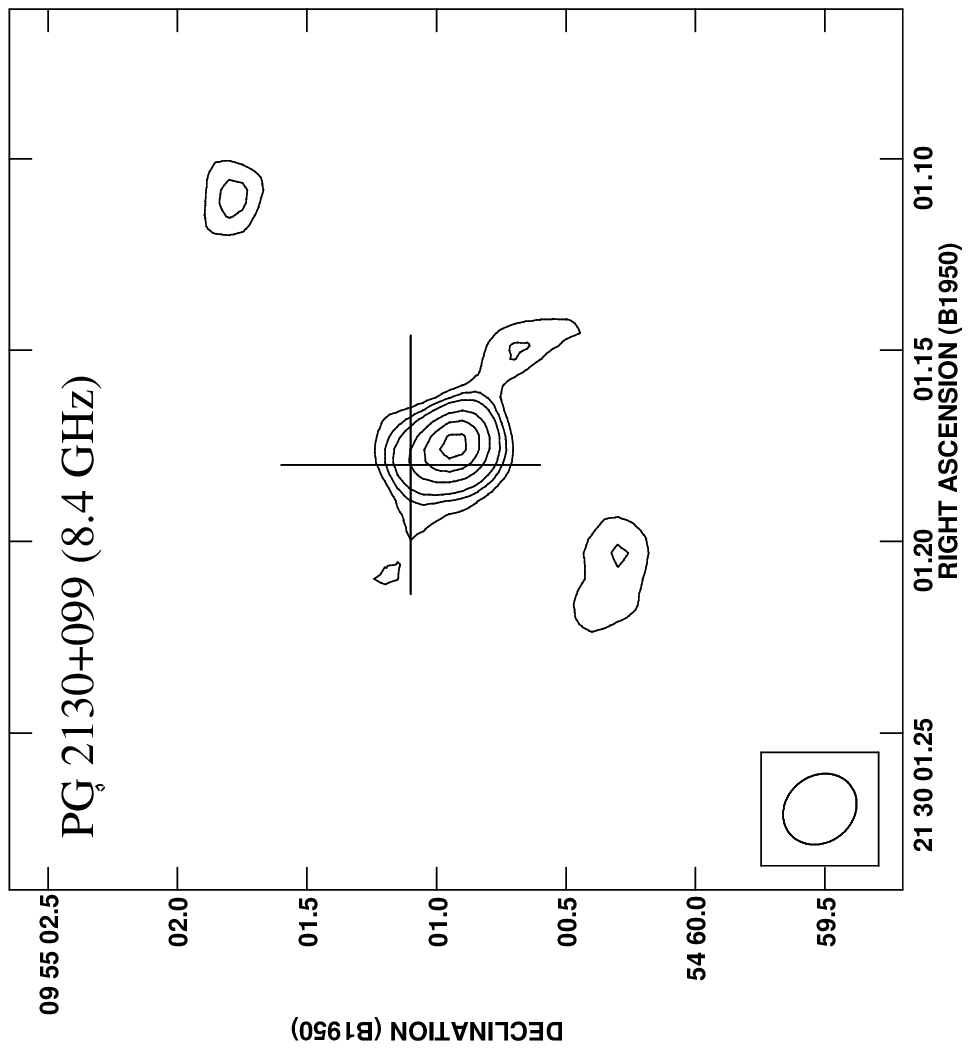}

\caption[]{Radio source structures at 8.4 GHz. From left to right:
PG~0157+001, PG~1612+261, PG~1700+518 \& PG~2130+099. Crosses mark the
position of the optical nucleus, scaled to represent a nominal
uncertainty of $\pm 0.5 ''$. The FWHM of the {\sc clean} beam is
indicated by the ellipse at the bottom left of each frame. Contour
levels are given in Table 3.}

\end{figure*}

\begin{table}
\caption{Contour levels for the radio maps in Figures 1 and 2
(Jy/beam). The base level is set to three times the rms noise in the
map.}

\centering
\begin{tabular}{cccl}
\hline
Object & Base & Contour levels ($\times$ base) \\ \hline
PG 0007+106 & $7.9\times 10^{-4}$ & -1, 1, 2, 4, 8, 16, 32,64, 128 \\
PG 0026+129 & $5.4\times 10^{-4}$ & -1, 1, 1.5, 2, 2.5, 3, 3.5 \\
PG 0052+251 & $3.2\times 10^{-4}$ & -1, 1, 1.5, 2, 2.5 \\
PG 0157+001 & $1.2\times 10^{-4}$ & 1, 2, 4, 8, 16 \\
PG 0921+525 & $5.0\times 10^{-4}$ & 1, 1.5, 2.2, 3.5, 5.5, 7, 8\\
PG 1612+261 & $1.1\times 10^{-4}$ & 1, 1.5, 2.5, 3.5, 4.5, 5.5\\
PG 1700+518 & $1.2\times 10^{-4}$ & 1, 2, 4, 8\\
PG 2130+099 & $9.1\times 10^{-5}$ & 1, 1.5, 2, 3, 4, 5\\
\hline
\end{tabular}
\end{table}

\subsection{Notes on individual objects}

\noindent
{\bf PG~0007+106.} This is the most radio-luminous object in the
current sample and is classed as a `radio-intermediate quasar' by
MRS93 and Falcke, Sherwood \& Patnaik (1996), rather than a true RQQ.
Since none of the three flux measurements in Table~4 were made
concurrently the precise values of the spectral indices must be
treated with caution, but the spectrum appears to be inverted. The
source is known to be variable at both radio and optical wavelengths
({\it eg} Schnopper et al. 1978, Ter\"{a}sranta et al. 1992). A comparison
of our December 1992 1.4-GHz data with the August 1991 map by Miller
shows that the flux of the core component has declined from 155.9~mJy
to 102.0~mJy over 16 months. The diffuse radio component 15$''$
(33~kpc) SW of the quasar in the 1.4-GHz map (Figure~1) was first
detected by Unger et al. (1987) and its flux density appears to have
remained constant at 8~mJy between 1983 and 1991.  This component does
not appear at 4.8 or 8.4~GHz, indicating a steep spectrum.

\noindent
{\bf PG~0026+129.} Due to problems on some of the shorter baselines
the noise in our 1.4-GHz map is relatively high and no radio
source is detected. However, Miller's deeper map of this object shows
a 2-mJy nuclear radio component with an extension $2''$ (6.5~kpc) to
the south (Figure~1). MRS93 failed to detect any radio emission at
4.8~GHz and the object was not observed at 8.4~GHz.

\noindent
{\bf PG~0050+124.} I~Zw~1 is one of the nearest objects in our sample
($z=0.061$) and is often classed as a Seyfert~1 rather than an RQQ
(see Section~5.5). As such it was observed with A-array at 8.4~GHz by
Kukula et al. (1995) as part of their radio survey of the CfA Seyfert
sample and we quote their radio position and flux measurement in
Table~4. The radio source is unresolved at all three frequencies,
suggesting a physical size of $\leq 0.38$~kpc. Using VLBI techniques
at 2295~MHz Roy et al. (1994) placed an upper limit of 0.1$''$ on the
angular size of the radio source, implying a brightness temperature
$T_{B} \geq 10^{5}$K. The simultaneous flux measurements at 1.4 and
4.8~GHz give the most reliable estimate of the spectral slope, which
appears to be steep ($\alpha^{1.4}_{4.8} = 0.7$). 

\noindent
{\bf PG~0052+251.} Our simultaneous flux measurements at 1.4 and
8.4~GHz in December 1992 show that PG~0052+251 has a relatively flat
radio spectrum ($\alpha^{1.4}_{8.4} = 0.4$). There is also some
evidence for variability in this source: the flux measured by MRS93 at
4.8~GHz nine years earlier is significantly lower than the value
predicted by the spectral index. Miller's slightly deeper 1.4-GHz
observations in August 1991 show marginal evidence for extended
emission to the south of the nucleus (Figure~1).

\noindent
{\bf PG~0157+001.} Although unresolved in our 1.4~GHz map, this object
appears as a $\sim2''$ (7~kpc) double source in both the 4.8-GHz map
of MRS93 and our new 8.4~GHz map (Figure 2). The two radio components
lie on either side of the optical nucleus, in the same PA as the
10-arcsec [O{\sc iii}] emission-line structure reported by Stockton \&
MacKenty (1987).

\noindent
{\bf PG~0804+761.} MRS93 and Kellermann et al. (1994) detect a $1''$
(2.5~kpc) double radio source at 4.8~GHz, but only the brighter,
northern component is visible in our 8.4-GHz image. From this we can
infer that the southern component either has a steep spectrum
($\alpha^{4.8}_{8.4} \geq 2.9$) or that it exhibits variability. Due
to the uncertainty in the optical position it is not clear which, if
any, of the components is associated with the quasar nucleus. At
1.4~GHz the double source cannot be resolved. Kellermann et al. (1994)
speculate that an 11-mJy elongated source 107$''$ to the north in
their 4.8-GHz D-array map might also be associated with the quasar,
but at a projected distance of 263~kpc we consider the possibility to
be unlikely and do not consider it further in the current work.

\noindent
{\bf PG~0921+525.} The 4.8-GHz images of MRS93 and Kellermann et
al. (1994) show a bright nuclear component from which a highly-curved
jet-like feature extends, initially to the NW. In our high-resolution
8.4-GHz image we detect only the central source, implying a steep
spectrum ($\alpha^{1.4}_{8.4} \geq 1.5$) for the components in the
`jet', but our 1.4-GHz map reveals a $3''$ (3~kpc) double source lying in
PA$\sim 20^{\circ}$ (Figure 1). All of the structure reported by MRS93
and Kellermann et al. lies within the brighter, southern component of
the 1.4-GHz double.

\noindent
{\bf PG~1012+008.} We derive a steep spectrum ($\alpha = 0.9$) for
this object between 1.4 and 8.4~GHz consistent with the results of
Barvainis et al. (1996) ($\alpha = 0.8$). However, we note that
Barvainis et al. also found evidence for a high-frequency turnover in
the spectrum at 15~GHz.

\noindent
{\bf PG~1116+215.}  We detect an unresolved ($\leq 1$~kpc) nuclear
component in this object, although we note that the D-array map by
Kellermann et al. (1994) also shows several radio sources $\sim$1
arcminute ($\sim230$~kpc) to the north and west.

\noindent
{\bf PG~1149$-$110.} At 8.4-GHz we detect an unresolved component
coincident with the optical quasar. The weak extensions visible in the
4.8-GHz image by MRS93 are almost certainly artifacts, as the authors
suggest.

\noindent
{\bf PG~1612+261.} This object possesses the most interesting radio
structure of all the RQQs in the current sample (Figure~2). At
4.8~GHz MRS93 found a bright, elongated radio component associated
with the optical nucleus, along with more diffuse emission extending
$\sim1''$ (3~kpc) in PA~$225^{\circ}$. Our 8.4-GHz map reveals
this structure in more detail: the nuclear component is resolved into
a 0.5-arcsec double source lying in the same PA as the extended
emission. The structure as a whole is approximately four times longer
than it is wide and arguably could be described as a one-sided jet
according to the criteria of Bridle \& Perley (1984) (the optical
nucleus lies at the eastern end of the radio structure).  Overall, the radio
source has a steep spectrum, consistent with the radio jets observed
in RLQs and in some nearby Seyfert nuclei.

\noindent
{\bf 1635+119.} Like PG~0007+106, the large radio luminosity and flat
spectrum of this unresolved point source places it in the category of
radio-intermediate quasars (RIQs). Gower \& Hutchings (1984) report a
second radio component $\sim53''$ ($\sim175$~kpc) south of the radio
core, which they claim to be a hotspot in a lobe associated with the
quasar.

\noindent
{\bf PG~1700+518.} Although unresolved at 1.4~GHz, this BAL quasar
appears as a $\sim1''$ (5.4~kpc) double source in both the 4.8~GHz
image of MRS93 and the current 8.4~GHz map (Figure~2). Both components
are slightly resolved; by comparing the {\it peak} flux densities at
4.8 and 8.4~GHz (2.2, 1.6~mJy~beam$^{-1}$ and 2.0, 1.3~mJy~beam$^{-1}$
respectively) we find that the eastern component has a slightly
flatter spectrum than its neighbour ($\alpha^{4.8}_{8.4} \sim
0.2,~0.35$ respectively). Barvainis et al. (1996) report a steep,
straight spectrum ($\alpha \sim 1$) in this object, but their
measurements were made with C-array and they would not have been able
to separate the emission from the two components. However, Kellermann
et al. (1994) do resolve the double structure at 15~GHz (with a
beamsize of $0.15''$) and report that the eastern source is brighter
at this frequency and consists of an unresolved component.  Although
the nominal position of the optical QSO (V\'{e}ron-Cetty \& V\'{e}ron
1993, marked by a cross in Figure 1) lies somewhere between the radio
components, a revised position (Hutchings, Neff \& Gower 1992) is
coincident with the eastern source to well within the positional
errors, suggesting that this is the synchrotron self-absorbed core of
the quasar.

\noindent
{\bf PG~2130+099.} This is a triple radio source with a bright central
component which is coincident with the optical nucleus, straddled by
two steep-spectrum lobes in PA$\sim135^{\circ}$ (Figure~2). The
overall extent of the structure is $2.5''$, equivalent to a projected
linear size of $\sim4$~kpc. The extension to the SW of the nuclear
component is probably the result of phase or amplitude errors which
could not be removed because the object was too weak for
self-calibration; it does not appear in the 4.8~GHz map of MRS93 even
though the two radio lobes are clearly visible. However, we note that
this feature is perpendicular to the triple source and lies in the
same PA as the major axis of the host galaxy (Taylor et al. 1996) and
therefore might be associated either with the galaxy itself or with
the structure which is responsible for defining the axis of the triple
source.  The A-array beam at 1.4 GHz is too large to resolve the
individual radio components but the source is slightly extended along
the axis of the triple structure. We find no evidence for any extensions
in the direction of the host galaxy major axis at this frequency.

\begin{table*}\label{radio}
\caption{Measured radio properties.  Radio positions for the
individual radio components are derived from our 8.4~GHz
A-configuration maps and are generally accurate to within 50~mas. For
objects which were either not observed or were undetected at 8.4~GHz
we give the position derived from the $L$-band (1.4~GHz) map (these
are marked (L) in column~3). Where an object with multiple radio
components is unresolved at 1.4~GHz we give the position of the peak
in the $L$-band map as well as the positions of the individual
components at 8.4~GHz.  The uncertainty in the spectral index values
is typically $\pm 0.02$.}
\centering
\begin{tabular}{rllrrrrrr}
\hline
Object & \multicolumn{2}{c}{Radio position (B1950)}& $S_{1.4GHz}$& $S_{4.8GHz}$& $S_{8.4GHz}$& \multicolumn{3}{c}{Spectral index} \\ 
            & \multicolumn{1}{c}{RA $h~m~s$}& \multicolumn{1}{c}{Dec $^{\circ}~'~''$} & mJy       & mJy      & mJy       & $\alpha^{1.4}_{4.8}$& $\alpha^{4.8}_{8.4}$& $\alpha^{1.4}_{8.4}$ \\ \hline
PG 0007+106 & 00 07 56.725 &+10 41 48.26     & $155.9\pm8.0^{\star}$  &$155.00^{*}    $  & $ 435\pm22$    &  0.0 & -1.8 &-0.6 \\
            & 00 07 55.82  &+10 41 40.3 (L)  & $  8.0\pm0.9^{\star}$  &        --     & $ \leq 4.5$    &   -- &   -- &  -- \\
PG 0026+129 & 00 26 38.05  &+12 59 29.6 (L)  & $  2.4\pm0.7^{\star}$  &$ \leq0.40^{*} $  &        --      &   -- &   -- &  -- \\
            & 00 26 38.05  &+12 59 27.9 (L)  & $  1.4\pm0.7^{\star}$  &$ \leq0.40^{*} $  &        --      &   -- &   -- &  -- \\
0046+112    & 00 46 55.435 &+11 12 05.98     & $  1.7\pm0.7$  &$  0.9\pm0.2   $  & $   0.5\pm0.2$ &  0.5 &  1.2 & 0.7 \\
PG 0050+124 & 00 50 57.77\dag&+12 25 19.21\dag&$  5.1\pm0.4$  &$  2.4\pm0.3   $  & $0.9\pm0.2$\dag&  0.7 &  1.8 & 1.0 \\
PG 0052+251 & 00 52 11.044 &+25 09 24.05     & $  1.1\pm0.5^{\star}$  &$  0.46^{*}    $  & $   0.7\pm0.1$ &  0.7 & -0.8 & 0.3 \\
0054+144    & 00 54 31.960 &+14 29 57.65     & $  1.8\pm0.4$  &$  0.73        $  & $   0.7\pm0.2$ &  0.7 &  0.2 & 0.6 \\
PG 0157+001 & 01 57 16.33  &+00 09 09.8 (L)  & $ 24.7\pm1.2$  &$  7.00^{*}$\ddag & $     4.9$\ddag&  1.0 &  0.6 & 0.7 \\
            & 01 57 16.336 &+00 09 09.82     &           --   &$  6.00^{*}    $  & $   4.0\pm0.2$ &   -- &  0.7 &  -- \\
            & 01 57 16.263 &+00 09 10.00     &           --   &$  1.00^{*}    $  & $   0.9\pm0.1$ &   -- &  0.2 &  -- \\
0244+194    & .............& .............    & $ \leq1.9   $  &$ \leq0.2      $  & $  \leq0.2   $ &   -- &   -- &  -- \\
0257+024    & 02 57 53.901 &+02 28 59.77     & $  5.5\pm0.7$  &$  3.4\pm0.2   $  & $   2.2\pm0.2$ &  0.4 &  0.8 & 0.5 \\
PG 0804+761 & 08 04 35.581 &+76 11 32.60     & $  1.3\pm0.8$  &$  1.00^{*}    $  & $   0.7\pm0.1$ &  0.2 &  0.8 & 0.4 \\
PG 0921+525 & 09 21 44.378 &+52 30 07.55     & $  5.8\pm0.6$  &$  1.90^{*}    $  & $   1.7\pm0.1$ &  0.9 &  0.2 & 0.7 \\
            & 09 21 44.43  &+52 30 10.0 (L)  & $  1.7\pm0.6$  &$ \sim0.5^{*}  $  & $  \leq0.12  $ &  1.0 &   -- &  -- \\
PG 0923+201 & .............& .............   & $ \leq0.4   $  &$ \leq0.4^{*}  $  & $  \leq0.2   $ &   -- &   -- &  -- \\
PG 0953+414 & .............& .............   & $ \leq2.6   $  &$ \leq0.20^{*} $  & $  \leq0.2   $ &   -- &  --  &  -- \\
PG 1012+008 & 10 12 20.797 &+00 48 33.55     & $  2.4\pm0.5$  &$  0.80^{*}    $  & $   0.5\pm0.1$ &  0.9 &  0.8 & 0.9 \\
PG 1116+215 & 11 16 30.122 &+21 35 43.00     & $  5.6\pm0.8$  &$  2.01^{*}    $  & $   1.4\pm0.1$ &  0.8 &  0.7 & 0.8 \\
PG 1149$-$110&11 49 30.326 &$-$11 05 42.75   & $  3.1\pm0.8$  &$  1.20^{*}    $  & $   1.3\pm0.1$ &  0.8 & -0.1 & 0.5 \\
PG 1211+143 & .............& .............   & $ \leq3.3   $  &$  \leq0.75^{*}$  &         --     &  --  &   -- &  -- \\
PG 1402+261 & 14 02 59.28  &+26 09 52.7 (L)  & $  1.1\pm0.4$  &$     0.45     $  &         --     &  0.7 &   -- &  -- \\
PG 1440+356 & 14 40 04.56  &+35 39 07.2 (L)  & $  3.9\pm1.2$  &$  0.78^{*}    $  &         --     &    --&   -- &  -- \\
1549+203    & .............& .............   & $ \leq1.0   $  &$  \leq0.2     $  & $  \leq0.2   $ &   -- &  --  &  -- \\
PG 1612+261 & 16 12 08.69  &+26 11 46.2 (L)  & $ 16.0\pm0.8$  &    --            &$3.3\pm0.2$\ddag&  --  &  --  & 0.9 \\
            & 16 12 08.727 &+26 11 46.60     &     --         &    --            & $   0.6\pm0.1$ &  --  &  --  &  -- \\
            & 16 12 08.704 &+26 11 46.45     &     --         &$  1.80^{*}    $  & $   0.7\pm0.1$ &  --  &  1.8 &  -- \\
            & 16 12 08.664 &+26 11 46.30     &     --         &    --            & $   0.3\pm0.1$ &  --  &  --  &  -- \\
PG 1613+658 & 16 13 36.236 &+65 50 37.75     & $  3.0\pm1.0$  &$  0.82^{*}    $  & $   1.4\pm0.1$ &  1.1 & -1.0 & 0.4 \\
1635+119    & 16 35 25.846 &+11 55 45.70     & $ 16.3\pm1.6$  &$ 15.3\pm0.8   $  & $  16.8\pm0.8$ & 0.005& -0.06&-0.02\\
PG 1700+518 & 17 00 13.32  &+51 53 36.4 (L)  & $ 23.2\pm1.2$  &$\sim3.8^{*}$\ddag&$4.3\pm0.1$\ddag&  1.5 & -0.2 & 0.9 \\
            & 17 00 13.286 &+51 53 36.20     &     --         &$  2.20^{*}    $  & $   2.6\pm0.2$ &   -- & -0.3 &  -- \\
            & 17 00 13.389 &+51 53 36.15     &     --         &$ \sim1.6^{*}  $  & $   1.7\pm0.1$ &   -- & -0.1 &  -- \\
PG 2130+099 & 21 30 01.16  &+09 55 01.1 (L)  & $ 58.2\pm2.9$  &$  1.67^{*}$\ddag &$0.9\pm0.1$\ddag&  2.9 &  1.1 & 1.8 \\
            & 21 30 01.176 &+09 55 00.95     &     --         &$  0.87^{*}    $  & $   0.6\pm0.1$ &      &  0.6 &     \\
            & 21 30 01.112 &+09 55 01.80     &     --         &$  0.4^{*}     $  & $   0.2\pm0.1$ &   -- &  1.2 &  -- \\
            & 21 30 01.203 &+09 55 01.30     &     --         &$  0.4^{*}     $  & $   0.1\pm0.1$ &   -- &  2.4 &  -- \\
2215$-$037  & .............& .............   & $ \leq11.2  $  &$ \leq0.3      $  & $  \leq0.4   $ &  --  &   -- &  -- \\
2344+184    & .............& .............   & $ \leq7.3   $  &$ \leq0.2      $  & $  \leq0.4   $ &  --  &   -- &  -- \\ \hline
\multicolumn{9}{l}{$^{\star}$ 1.4~GHz measurements taken from the August 1991 observations by Miller. Additional 1.4~GHz measurements from} \\ 
\multicolumn{9}{l}{December 1992: PG~0007+106, $102\pm5$~mJy; PG~0026+129, $\leq 2.75$~mJy; PG~0052+251, $1.4\pm 0.6$~mJy.} \\
\multicolumn{9}{l}{$^{*}$ 4.8~GHz {\it peak} flux densities taken from MRS93. Uncertainties are typically 0.2~mJy.} \\
\multicolumn{9}{l}{\dag~8.4~GHz measurement taken from Kukula et al. (1995).} \\
\multicolumn{9}{l}{\ddag~Total flux density (sum of the individual components).} \\

\end{tabular}
\end{table*}

\begin{table*}
\label{derived}
\caption{Derived radio properties. Maximum angular extent is defined as either
the distance between the peaks of the most widely separated radio
components in resolved sources or the FWHM of the restoring beam in
unresolved sources. Radio morphologies are defined as
follows: P $=$ point source, D $=$ double, T $=$ triple, CL $=$
core-lobe, and CJ $=$ core-jet. Two objects were unresolved in the
current observations but are known to possess arcsec-scale radio
structure from previous studies: PG~0804+761 (MRS93) and 1635+119
(Gower \& Hutchings 1984). K-corrections were performed using the radio
spectral indices, $\alpha^{1.4}_{8.4}$, given in column~9 of Table~4
or, where such information was not available, by assuming a spectral
index of 0.7.  Entries marked with a \ddag~ give the {\it total} radio
luminosity in the A-array map, obtained by combining the fluxes of the
individual radio components.  Brightness temperatures have been
calculated at 8.4~GHz for the individual radio components using both
the nominal $0.24''$ A-array beam as an upper limit to the angular
size, and the deconvolved size given by the {\sc aips} task {\sc
jmfit} (in many cases this is an upper limit only).}

\centering
\begin{tabular}{rccrcrrrrr}
\hline
\multicolumn{1}{c}{Object} & $M_{V}$ &\multicolumn{2}{c}{Max. extent} & \multicolumn{1}{c}{Radio} &\multicolumn{3}{c}{log$_{10}$(L$_{radio}$) (W~Hz$^{-1}$)}& \multicolumn{2}{c}{Brightness temperature (K)}  \\ 
       &         & $''$ & kpc & \multicolumn{1}{c}{morph.} & 1.4 GHz & 4.8 GHz & 8.4 GHz & $0.24''$ beam  & Deconvolved \\ \hline
PG 0007+106 &$-$23.24 &  15.0 &  33.4& CL   &   24.49 &   24.68 &   25.12 &  $> 1.05 \times 10^{5}$ &  $> 1.47 \times 10^{6}$\\
            &         &       &      &      &   23.39 &   --    &$<$23.14 &         --              &         --             \\     
PG 0026+129 &$-$24.24 &   2.0 &   6.5& D    &   23.32 &$<$22.55 &   --    &         --              &         --             \\
            &         &       &      &      &   23.10 &$<$22.55 &   --    &         --              &         --             \\   
0046+112    &$-$24.03 &$<$0.24&$<$1.3& P    &   23.81 &   23.52 &   23.23 &  $> 1.13 \times 10^{2}$ &  $> 1.63 \times 10^{4}$\\
PG 0050+124 &$-$23.82 &$<$0.24&$<$0.4& P    &   22.96 &   22.59 &   22.14 &  $> 2.17 \times 10^{2}$ &  $> 1.25 \times 10^{3}$\\
PG 0052+251 &$-$23.93 &$\sim$1&$\sim$3.5&CJ?&   23.15 &   22.65 &   22.84 &  $> 1.73 \times 10^{2}$ &  $> 6.93 \times 10^{2}$\\
0054+144    &$-$24.34 &$<$0.24&$<$0.9& P    &   23.37 &   22.97 &   22.92 &  $> 1.56 \times 10^{2}$ &  $> 9.01 \times 10^{6}$\\
PG 0157+001 &$-$24.27 &   2.0 &   7.3& D    &   24.48 &23.94\ddag&23.77\ddag&       --              &         --             \\
            &         &       &      &      &   --    &   23.87 &   23.69 &  $> 9.53 \times 10^{2}$ &  $> 3.43 \times 10^{4}$\\
            &         &       &      &      &   --    &   23.08 &   23.02 &  $> 2.12 \times 10^{2}$ &  $> 3.05 \times 10^{6}$\\
0244+194    &$-$23.46 & --    & --   & --   &$<$23.47 &$<$22.53 &$<$22.44 &         --              &         --             \\
0257+024    &$-$23.09 &$<$0.24&$<$0.7& P    &   23.50 &   23.29 &   23.10 &  $> 5.34 \times 10^{2}$ &  $  1.44 \times 10^{4}$\\
PG 0804+761 &$-$23.74 &   1.0 &   2.5& D    &   22.75 &   22.63 &   22.44 &  $> 1.56 \times 10^{2}$ &  $> 2.25 \times 10^{6}$\\
PG 0921+525 &$-$20.61 &   3.0 &   2.9& D    &   22.49 &   22.00 &   21.95 &  $> 4.07 \times 10^{2}$ &  $> 1.50 \times 10^{3}$\\
            &         &       &      &      &   21.95 &$\sim$21.42&$<$20.80&        --              &         --             \\
PG 0923+201 &$-$24.45 & --    & --   & --   &$<$22.84 &$<$22.76 &$<$22.38 &         --              &         --             \\
PG 0953+414 &$-$25.23 & --    & --   & --   &$<$22.97 &$<$22.79 &$<$22.72 &         --              &         --             \\
PG 1012+008 &$-$24.38 &$<$0.24&$<$1.0& P    &   23.58 &   23.10 &   22.90 &  $> 1.20 \times 10^{2}$ &  $> 4.33 \times 10^{3}$\\
PG 1116+215 &$-$25.09 &$<$0.24&$<$0.9& P    &   23.90 &   23.46 &   23.29 &  $> 3.27 \times 10^{2}$ &  $> 1.18 \times 10^{4}$\\
PG 1149$-$110&$-$21.88&$<$0.24&$<$0.3& P    &   22.50 &   22.09 &   22.11 &  $> 3.03 \times 10^{2}$ &  $> 6.99 \times 10^{3}$\\
PG 1211+143 &$-$23.91 & --    & --   & --   &$<$23.02 &$<$22.38 &   --    &         --              &         --             \\
PG 1402+261 &$-$24.14 &$<$0.4 &$<$1.5& P    &   23.10 &   22.73 &   --    &         --              &         --             \\
PG 1440+356 &$-$23.74 &$<$0.4 &$<$0.8& P    &   23.01 &   22.31 &   --    &         --              &         --             \\
1549+203    &$-$24.38 & --    & --   & --   &$<$23.50 &$<$22.82 &$<$22.71 &         --              &         --             \\
PG 1612+261 &$-$24.07 &   1.0 &   3.1& CJ   &   24.10 &      -- &23.41\ddag&        --              &         --             \\
            &         &       &      &      &   --    &      -- &   22.67 &  $> 1.52 \times 10^{2}$ &  $> 8.73 \times 10^{6}$\\
            &         &       &      &      &   --    &   23.15 &   22.73 &  $> 1.61 \times 10^{2}$ &  $> 9.29 \times 10^{6}$\\
            &         &       &      &      &   --    &      -- &   22.37 &  $> 8.18 \times 10^{1}$ &  $> 4.71 \times 10^{6}$\\ 
PG 1613+658 &$-$23.95 &$<$0.24&$<$0.7& P    &   23.33 &   22.76 &   23.00 &  $> 3.42 \times 10^{2}$ &  $  1.31 \times 10^{8}$\\
1635+119    &$-$23.21 &  53.0 & 175.0& CL   &   24.15 &   24.12 &   24.16 &  $> 4.04 \times 10^{3}$ &  $> 1.92 \times 10^{6}$\\
PG 1700+518 &$-$25.77 &   1.0 &   5.4& D    &   24.97 &24.18\ddag&24.23\ddag&       --              &         --             \\
            &         &       &      &      &   --    &   23.95 &   24.02 &  $> 6.21 \times 10^{2}$ &  $> 3.97 \times 10^{6}$\\
            &         &       &      &      &   --    &$\sim$23.79& 23.82 &  $> 4.14 \times 10^{2}$ &  $> 2.65 \times 10^{6}$\\
PG 2130+099 &$-$23.25 &   2.5 &   4.1& T    &   24.05 &22.48\ddag&22.24\ddag&       --              &         --             \\
            &         &       &      &      &   --    &   22.22 &   22.07 &  $> 1.47 \times 10^{2}$ &  $> 8.45 \times 10^{6}$\\
            &         &       &      &      &   --    &   21.84 &   21.57 &  $> 4.81 \times 10^{1}$ &  $> 2.77 \times 10^{6}$\\
            &         &       &      &      &   --    &   21.84 &   21.27 &  $> 2.41 \times 10^{1}$ &  $> 1.39 \times 10^{6}$\\
2215$-$037  &$-$23.60 & --    & --   & --   &$<$24.58 &$<$22.98 &$<$23.10 &         --              &         --             \\
2344+184    &$-$23.69 & --    & --   & --   &$<$23.84 &$<$22.28 &$<$22.58 &         --              &         --             \\ \hline
\multicolumn{7}{l}{\ddag~Total radio luminosity.} \\
						       
\end{tabular}
\end{table*}

\section{The origin of the radio emission in RQQs}

In this section we describe various models to explain the compact
nuclear radio emission in RQQs and discuss the ways in which they can
be distinguished observationally.

Kellermann et al. (1989), in their radio survey of the BQS quasars,
compare the total flux at 4.8~GHz (measured with D-array with a
resolution of $18''$) with the flux of the central {\sc clean}
component in their high-resolution ($0.24''$ FWHM) A-array
maps. Taking the average value of this ratio for the radio-quiet BQS
objects reveals that the central component accounts for 72\% of the
total radio emission from the quasar. In other words almost three
quarters of the total radio luminosity of an RQQ comes from a very
small ($\ll1$~arcsec) region in the nucleus of the host galaxy.

The current observations were all made with the high-resolution
A-configuration of the VLA and are therefore more sensitive to the
compact nuclear radio sources than to diffuse emission from the rest
of the host galaxy.  In Table~6 we list the total and compact radio
fluxes of all the RQQs in our sample at both 1.4 and 4.8 GHz. In this
case the `compact' flux refers to all of the emission detected in our
A-array maps with FWHM beam sizes of 1.4$''$ and 0.4$''$ at 1.4~GHz
and 4.8~GHz respectively. In the case of PG~1612+261 we do not have a
measured flux at 4.8~GHz for the extended structure in the A-array map
(MRS93 give only the peak value) so we have extrapolated from the flux
at 1.4~GHz using the measured spectral index of $\alpha^{1.4}_{8.4}=
0.9$.  The `total' fluxes are D-array measurements taken from the
literature or downloaded from the Internet. At 1.4~GHz we used frames
taken from the NRAO/VLA Sky Survey (NVSS; Condon et al. 1996) with a
beam size of $40''$ (38-220~kpc for $z =$ 0.035-0.3). The 4.8-GHz
fluxes are from Kellermann et al. (1989) and have a beam size of
$18''$ (17-99~kpc for $z =$ 0.035-0.3). The ratio, $\rm f_{c}$, of the
compact to the total radio emission is given at each frequency. The
average values for $\rm f_{c}$ are 0.67 and 0.64 at 1.4 and 4.8~GHz
respectively - demonstrating that the majority of the total radio
emission from the RQQ comes from the compact features in the nucleus
rather than the body of the host galaxy. The distribution of $\rm
f_{c}$ shows this more clearly (Figure~3). It should be noted that no
attempt has been made to correct for any variability in the intrinsic
radio luminosity of the quasar between the dates of the D- and A-array
measurements.

\begin{table}
\caption[]{Total (galaxy plus nucleus) and compact radio fluxes. Total
radio fluxes at 4.8~GHz are taken from Kellermann et al. (1989):
D-array, 18$''$ beam, 3-sigma uncertainty $=0.2$~mJy, observations
taken between November 1982 and May 1983). Total fluxes at 1.4~GHz are
from the NVSS: D-array, 40$''$ beam, 3-sigma uncertainty $=1.2$~mJy,
observations taken in October/November 1993 (for Right Ascensions
23~hrs through to 14.5~hrs) and April/May 1995 (for remaining
RAs). Compact fluxes were obtained by summing the A-array fluxes of
the individual radio components in Table~4. The quantity f$_{\rm c}$
is the ratio of the two, {\it ie} the fraction of the total flux
density which is contained in the compact nuclear components. The
4.8-GHz A-array flux of PG~1612+261 was estimated from the flux at
1.4~GHz, using the measured spectral index of 0.9.}

\begin{tabular}{rrrcrrc}
\hline
             &\multicolumn{3}{c}{1.4-GHz flux (mJy)} &\multicolumn{3}{c}{4.8-GHz flux (mJy)} \\
             &\multicolumn{1}{c}{D}&\multicolumn{1}{c}{A}& &\multicolumn{1}{c}{D}&\multicolumn{1}{c}{A}&  \\
Source       &  Array &  Array  & f$_{\rm c}$&Array  & Array    & f$_{\rm c}$\\
\hline
PG 0007+106  &  100.1 &   102.0 &   1.02 &   321.00  &   155.00 & 0.48 \\
PG 0026+129  &    7.6 &     3.8 &   0.50 &     5.10  &  $<$0.40 & --   \\
0046+112     & $<$1.2 &     1.7 &   --   &     --    &     0.9  & --   \\
PG 0050+124  &    9.0 &     5.1 &   0.57 &     2.60  &     2.4  & 0.92 \\
PG 0052+251  &    --  &     1.4 &   --   &     0.74  &     0.46 & 0.62 \\
0054+144     &    2.4 &     1.8 &   0.75 &     --    &     0.73 & --   \\
PG 0157+001  &   27.4 &    24.7 &   0.90 &     8.00  &     7.00 & 0.88 \\
0244+194     & $<$1.1 &  $<$1.9 &   --   &     --    &  $<$0.2  & --   \\
0257+024     &    5.7 &     5.5 &   0.96 &     --    &     3.4  & --   \\
PG 0804+761  &    3.4 &     1.3 &   0.38 &     2.38  &     1.00 & 0.42 \\
PG 0921+525  &   10.4 &     7.5 &   0.72 &     3.80  &     2.4  & 0.63 \\
PG 0923+201  & $<$1.1 &  $<$0.4 &   --   &     0.25  &  $<$0.36 & --   \\
PG 0953+414  &    4.7 &  $<$2.5 &   --   &     1.90  &  $<$0.20 & --   \\
PG 1012+008  &    --  &     2.4 &   --   &     --    &     0.80 & --   \\
PG 1116+215  &    --  &     5.6 &   --   &     2.80  &     2.01 & 0.72 \\
PG 1149$-$110&   10.5 &     3.1 &   0.30 &     2.60  &     1.20 & 0.46 \\
PG 1211+143  &    --  &  $<$3.3 &   --   &   157.00  &  $<$0.75 & --   \\
PG 1402+261  &   13.8 &     1.1 &   0.08 &     0.62  &     0.45 & 0.73 \\
PG 1440+356  &    5.1 &     3.9 &   0.76 &     1.66  &     0.78 & 0.45 \\
1549+203     &    --  &  $<$1.0 &   --   &     --    &  $<$0.22 & --   \\
PG 1612+261  &   18.6 &    16.0 &   0.86 &     5.07  &     5.30 & 1.04 \\
PG 1613+658  &    4.2 &     3.0 &   0.71 &     3.03  &     0.82 & 0.27 \\
1635+119     &    --  &    16.3 &   --   &     --    &    15.3  & --   \\
PG 1700+518  &    --  &    23.2 &   --   &     7.20  &     3.8  & 0.53 \\
PG 2130+099  &    --  &    58.2 &   --   &     2.05  &     1.67 & 0.81 \\
2215-037     &    --  & $<$11.1 &   --   &     --    &  $<$0.28 & --   \\
2344+184     & $<$1.0 &  $<$7.3 &   --   &     --    &  $<$0.20 & --   \\
\hline	   
\end{tabular}
\end{table}

\begin{figure}
\vspace {10.5truecm}
 
\includegraphics{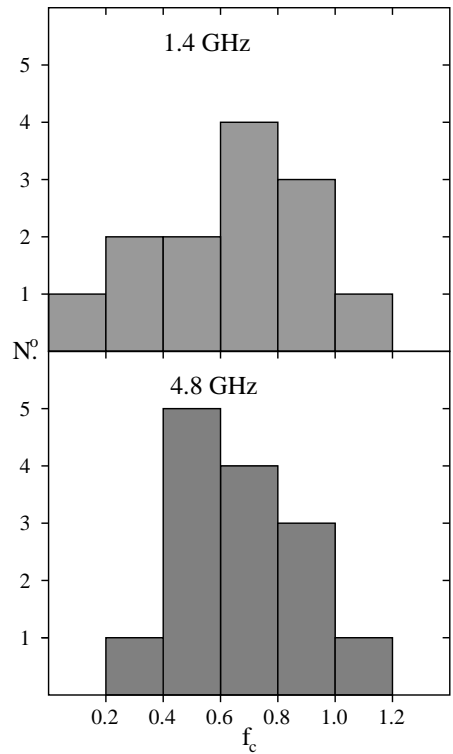}

\caption[]{Distribution of the fraction, $\rm f_{c}$, of the total
radio emission of the quasar (measured with D-array) which originates
in the compact nuclear radio components ({\it ie} all the flux
detected with A-array).}

\end{figure}

Clearly the nuclear emission contributes a significant, often
dominant, fraction of the total radio luminosity of the RQQ. Whatever
process is responsible for these compact components must be capable of
producing large amounts of radio emission - comparable to the amount
produced by the rest of the galaxy - within a region less than a
kiloparsec across.

The radio sources in radio-loud AGN consist of both large-scale,
highly-collimated structures and very compact features, sometimes with
flat or inverted radio spectra. These latter often exhibit high
brightness temperatures, extreme variability and apparently
superluminal motions - phenomena for which the only plausible
explanation seems to be a relativistic jet produced by a `central
engine' (Begelman, Blandford \& Rees 1984). Thus, for radio-loud
objects it seems indisputable that the ultimate source of the radio
emission is a compact central engine, or `monster'; conventionally a
supermassive black hole.

The origin of the much weaker radio emission in radio-quiet AGN is far
less certain, and doubts remain as to whether the emission is directly
associated with the AGN itself. Kellermann et al. (1994) found that
the 5-GHz radio luminosities of the radio-quiet BQS objects were much
larger than the typical values for spiral or elliptical galaxies, but
only a handful of RQQs (3/92) exhibited linear kpc-scale emission.
Sopp \& Alexander (1991) point to the radio/FIR correlation in RQQs as
evidence that the source of the `extra' radio emission is a
circumnuclear starburst. In a more radical approach, Terlevich et
al. (1992) have attempted to explain the entire phenomenon of
radio-quiet activity, including the radio emission, in terms of a
dense, {\it nuclear} starburst rather than a central engine.

There are several differences between the radio emission produced by
regions of enhanced starformation and that directly associated with
the central engine of an AGN (Condon 1992), and these can, in
principle, be used to distinguish between the two mechanisms.

In starburst regions we observe both non-thermal synchrotron radiation
and thermal free-free emission, the former from radio supernovae (RSN)
and supernova remnants (SNR) and the latter from H{\sc ii}
regions. The radio emission can be distributed over a large volume but
will be constrained to lie roughly within the isophotes of the host
galaxy (though some of the plasma might diffuse out of the galaxy
along local density gradients). With very high resolution numerous
small radio sources might be resolved, corresponding to individual
supernova remnants ({\it eg} those in M82, Muxlow et al. 1994).  The
brightness temperature of starburst-related radio emission should not
exceed $T_{B}\sim 10^{5}$K and the emission becomes optically thick,
with a flat spectrum, for $T_{B} > 10^{4}$K. The explosion of
RSN can lead to random variations in the radio emission.

By contrast, radio-loud AGN produce non-thermal synchrotron emission
from highly-collimated jets of plasma which contain compact knots and
terminate in diffuse radio lobes. The knots are typically several
orders of magnitude more luminous than individual supernova
remnants. Extremely high apparent brightness temperatures are
observed, but the emission remains optically thin with a steep
spectrum until synchrotron self-absorption becomes significant at
$T_{B} \sim 10^{10}$K. Flat-spectrum, self-absorbed radio cores are
frequently associated with the optical nucleus of the AGN.  When a
relativistic jet is closely aligned with the line of sight the radio
emission is boosted in the direction of the observer and the source
will be highly variable with a flat radio spectrum and apparent
brightness temperatures of up to $10^{15}$~K. In such cases apparent
superluminal motions are sometimes also observed.

Strong evidence is now emerging that at least some RQQs contain radio
sources which are fundamentally similar to those in RLQs, albeit
several orders of magnitude smaller and less powerful.  Barvainis et
al. (1996) found evidence for a flat-spectrum radio component in
40$\%$ of their 23 RQQs and several of these sources also showed the
variability and high brightness temperatures characteristic of the
radio emission in RLQs. They conclude that at least some RQQs contain
a partially opaque synchrotron core which is associated with a compact
central engine.  

The presence of relativistic jets in at least some RQQs has been
inferred by MRS93 from the correlation between radio and
[OIII]$\lambda$5007 luminosities (though in many cases an
extra-nuclear starburst is also required in order to explain
correlations between non-nuclear radio and [OIII] emission). Also,
several authors have argued that the simplest interpretation of the
so-called `radio-intermediate quasars' (RIQs) is that they are objects
which have been boosted out of the radio-quiet population by
relativistic beaming effects (MRS93, Falcke, Malkan \& Biermann 1995,
Falcke, Sherwood \& Patnaik 1996). This in turn implies the existence
of misaligned radio jets in at least some RQQs.

Finally, high-resolution mapping with VLBI has revealed extremely
compact (milliarcsecond-scale), high $T_{B}$ radio components in a
handful of RQQs ({\it eg} Hutchings \& Neff 1992) including the object
E1821+643 which contains both a compact, flat spectrum, pc-scale core
and a steep-spectrum, 100-pc-scale radio jet (Papadopoulos et
al. 1995, Blundell \& Lacy 1995, Blundell et al. 1996).

\section{Implications of the current data}

Our observations have enabled us to determine the radio luminosities,
spectral indices and, in some cases, morphologies of the RQQs in our
sample. Using this information we can now attempt to place some
constraints on the processes responsible for the compact radio
emission in RQQs and to compare their radio properties with those of
other types of AGN. 

\subsection{Supernovae and supernova remnants}

Both supernova remnants (SNRs) and recently exploded supernovae can be
strong sources of radio emission, with the latter up to 300 times as
powerful as Cas A.  Ulvestad (1982) used the surface brightness --
diameter ($\Sigma - D$) relation for galactic SNRs to estimate the
supernova rate required to produce the observed radio luminosity of
nearby Seyfert nuclei. In the majority of cases an unfeasibly large
supernova rate is implied.  The number of massive stars required to
maintain such a rate would have an energy output at optical
wavelengths far in excess of the observed luminosities of Seyferts.

However, this method assumes that all the energy of the supernova is
released during the adiabatic (Sedov) phase and thus ignores emission
from remnants older than $\sim 2 \times 10^{4}$ years.  As a
result, the $\Sigma - D$ relation overestimates the supernova
rate of our own Galaxy and predicts flatter radio spectral indices for
nearby spirals than are actually observed. Condon \& Yin (1990) argue
that the ratio of our Galaxy's non-thermal radio emission to its
supernova rate gives a more reliable means of estimating supernova
rates:
\begin{equation}
L_{\nu} ({\rm W Hz^{-1}}) \simeq 1.3 \times 10^{23} (\nu_{GHz})^{-\alpha} \gamma_{SN} ({\rm yr^{-1}}) 
\end{equation}
where $\alpha$ is the radio spectral index and $\gamma$ is the
supernova rate per year. Applying this formula to our 8.4-GHz A-array
fluxes gives supernova rates ranging from $2\times10^{-2}$ to 50 per
year to power the compact radio components in RQQs. Although the lower
figure is comparable to the supernova rate in our Galaxy, the upper
value is $\sim 30$ times larger than the typical supernova rates in
nearby starburst galaxies. In the compact radio sources in RQQs all of
this activity must take place within a region which is typically less
than a kiloparsec across. Thus it seems unlikely that supernovae and
their remnants can account for the non-thermal emission, except in the
faintest {\it and} largest nuclear radio sources.

\subsection{Brightness temperatures}

Table 5 lists the 8.4-GHz brightness temperatures for all the detected
radio components.  Since all of the components are essentially
unresolved the FWHM of the A-array beam has been used to give an upper
limit on the angular size of 0.24$''$ - the corresponding value of
$T_{B}$ is therefore a {\it lower limit} only. 

Using this method gives typical lower limits on $T_{B}$ in the range
$\sim 10^{2}$ to $\sim 10^{3}$K. Such brightness temperatures could
easily be produced by either the starburst or the AGN mechanism.
However, deconvolving the radio component from the A-array beam using
the {\sc aips} task {\sc jmfit} allows us to derive a much more
stringent limit on the angular size. The reliability of the
deconvolution algorithm is highly dependent on the signal-to-noise
ratio in the radio map. For our 8.4-GHz snapshot observations this
ratio is relatively low (peak flux densities are typically $\sim
1$~mJy~beam$^{-1}$, with $3\sigma$ noise levels of
$0.1$~mJy~beam$^{-1}$) and the deconvolved sizes must therefore be
treated with caution.  However, using this method we find that in the
majority of objects the deconvolved radio source is extremely compact,
leading to values of $T_{B}$ which are typically $\sim10^{4}$ to
$\sim10^{6}$K (and as high as $10^{8}$K in one case). For $T_{B} >
10^{4}$K we would expect any starburst related radio emission to
become optically thick with a flat spectrum, whereas we find that in
the majority of cases the radio emission between 1.4 and 8.4~GHz has a
steep spectrum. 

The VLBI measurements necessary to confirm these values of $T_{B}$
have so far been carried out for only a handful of RQQs ({\it eg}
Hutchings \& Neff 1992, Roy et al. 1994, Blundell et al. 1996), but
temperatures in excess of $10^{5}$K appear to be common.  This
constitutes strong evidence that the primary source of the compact
radio emission in RQQs is non-stellar - {\it ie} a central engine
rather than a starburst.

\subsection{Radio structures}

If the radio emission in RQQs can be shown to take the form of
highly-collimated jets as it does in RLQs and Radio Galaxies then we
can confidently ascribe it to the central engine of the RQQ and rule
out starburst regions as the primary source of the radio
plasma. Although the jet collimation mechanism in radio-loud AGN is
not understood there is no viable mechanism by which an ensemble of
supernovae could produce continuous, highly-collimated outflows.

At 8.4~GHz the angular diameter (FWHM) of the A-array beam is
0.24$''$, corresponding to a spatial resolution of $\sim 0.3$~kpc to
$\sim 1.3$~kpc over the redshift range of our RQQ sample.
Arcsecond-scale radio structures have been detected in 9 of our
objects (10 if we accept the marginal evidence for extended emission
in PG~0052+251), although none of these can unambiguously be
identified as jets (Table~5).  However, in each case the radio morphology
conforms to the pattern of double, triple and linear sources familiar
from maps of radio-loud quasars - where it is invariably associated
with powerful jets from the AGN. Conceivably a starburst taking place
in a nuclear bar might be able to produce symmetrical or elongated
radio morphologies, but not highly collimated jet-like structures.
Forthcoming MERLIN observations of our RQQs, with a five-fold
improvement in angular resolution, should enable us to test this
possibility.

The emitting regions in the resolved RQQs are all at least 2~kpc
across and are thus consistent with the largest radio structures
observed in nearby Seyfert galaxies - objects in which detailed
studies have also found evidence for radio jets (see Section
5.4). However, in the unresolved RQQs the upper limits on the physical
sizes are typically $\sim1$~kpc or less. This is also consistent with
radio studies of Seyfert nuclei, in which $\sim50\%$ of objects appear
to possess only a compact ($\ll 100$~pc) nuclear radio source, with no
evidence for extended structure ({\it eg} Kukula et al. 1995).

\subsection{Comparison with Seyfert nuclei}

There are many similarities between the radio properties of Seyfert
galaxies and those of the RQQs in the present survey.  In both cases
steep-spectrum, non-thermal emission and high brightness temperatures
are observed.  Although most of the RQQs are unresolved, the upper
limits on their sizes place their radio sources in the same size
regime as those of Seyferts. Where the radio source can be resolved we
find double, triple and more complex radio structures, all highly
reminiscent of early, low-resolution radio maps of Seyfert nuclei.

These similarities are unsurprising. In terms of spectroscopic
characteristics there is no real difference between RQQs and the
nuclei of Type 1 Seyfert galaxies - the distinction is purely one of
optical luminosity. Historically, the classification of a particular
object has depended on the method of discovery and is essentially a
result of whether the host galaxy is easily detectable from the
ground.  In order to formalise the classification system Schmidt \&
Green (1983) proposed an arbitrary distinction in terms of absolute
magnitudes: objects with $M_{B} < -23$ are RQQs, whilst anything with
$M_{B} > -23$ is deemed a Seyfert galaxy. Although this definition has
been generally adopted, low-redshift ($z \leq 0.1$) objects with
$M_{B} < -23$ are still frequently referred to as Seyfert nuclei,
particularly when the surrounding galaxy is clearly visible.

The relatively small distances to many Seyferts mean that the radio
structures of their nuclei can be investigated on much smaller spatial
scales than is possible for RQQs, providing a greater insight into the
processes responsible for the radio emission.  Typical flux densities
of a few milliJanskys are measured, corresponding to radio
luminosities L$_{5GHz}< 10^{24}$W~Hz$^{-1}$.  Although the total radio
emission of Seyfert galaxies includes contributions from the galaxy
disc and from starburst regions, in many objects there is a strong
radio source which is directly associated with the active nucleus.
This source is generally confined to the central regions of the host
galaxy but can exhibit structures on a range of scales from
$\sim$10~kpc down to $\sim$1~pc at the limit of current resolving
power. Radio structures are frequently linear or `jet-like', {\it ie}
elongated or consisting of a series of aligned radio knots. Spectra
are usually steep ($\alpha \sim0.7$) but flatter spectra ($\alpha \leq
0.3$) are sometimes found in compact radio components. There can be
little doubt that these structures represent radio jets, similar to,
but much weaker than, those in radio-loud AGN, and that they are
capable of contributing a significant fraction of the total radio
luminosity of the galaxy, in some cases becoming the {\it dominant}
source of radio emission ({\it eg} de Bruyn \& Wilson 1976, Ulvestad \&
Wilson 1984a,b, Kukula et al. 1995).

As we demonstrate in the next subsection, the distribution of radio
luminosities in the RQQs is consistent with them being more radio
powerful, as well as more optically luminous, versions of Seyfert 1
nuclei. The VLBA observations of E1821+643 by Blundell et al. (1996)
have already demonstrated the existence of Seyfert-like radio
structures in one RQQ, and there appears to be no reason why this
should not also be the case in many other objects. Observations with
high angular resolution, such as our own forthcoming MERLIN survey,
will help to clarify this issue. 

\subsection{The radio-optical relation in radio-quiet AGN}

\begin{figure*}
\vspace {11.5truecm}
 
\includegraphics{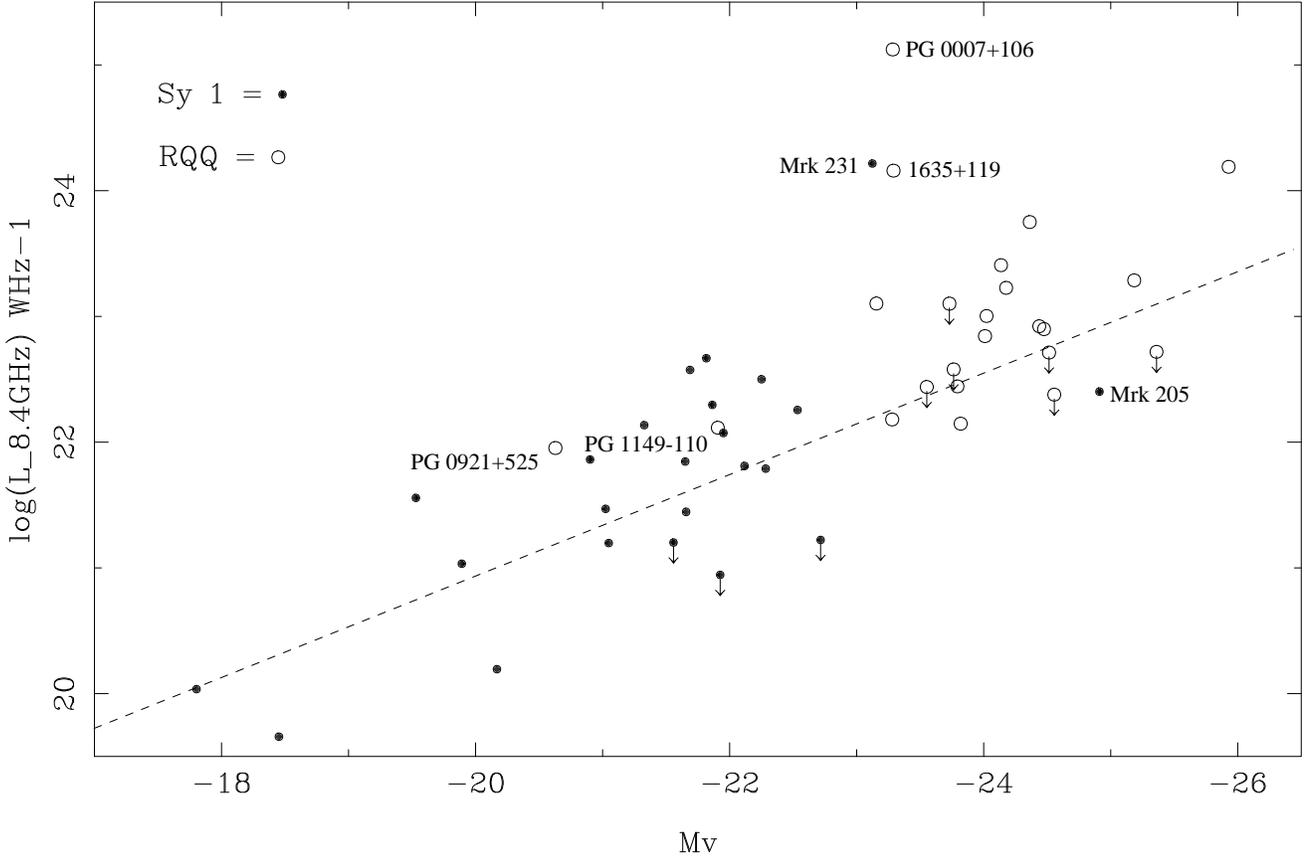}

\caption[]{8.4-GHz radio luminosity as a function of $M_{V}$ for the
RQQs in the current sample (open circles) and Type 1 Seyferts from the
CfA sample of Huchra \& Burg (1992) (filled circles). Upper limits are
denoted by open and filled triangles respectively. The radio
luminosities of the Seyfert 1s were derived from the C-array
observations of Kukula et al. (1995). The dotted line shows the best
fit to the data taking into account the radio upper limits.}

\end{figure*}

The smooth increase in optical luminosities from the faintest Seyferts
through into the RQQ population impelled us to look for a similar
effect in the distribution of radio luminosities. The existence of a
correlation between the radio and optical emission in radio-quiet AGN
would constitute further evidence for a direct link between the
central engine of the quasar, which produces the optical continuum,
and the processes responsible for the nuclear radio emission.

Several authors have found evidence for radio-optical luminosity
correlations in both radio-loud (Rawlings \& Saunders 1991, Serjeant
et al. 1997) and radio-quiet quasars (Stocke et al. 1992, Lonsdale,
Smith \& Lonsdale 1995). However, the existing evidence is not
conclusive and the situation is complicated by the difficulty of
separating optical luminosity from redshift in the major, flux-limited
quasar samples.

In Figure~4 we plot the 8.4-GHz radio luminosities of the RQQs in our
current sample as a function of absolute magnitude, $M_{V}$. Also
shown on this plot are the Type 1 Seyferts from the CfA sample (Huchra
\& Burg 1992). The radio luminosities of the Seyfert 1s are derived
from the C-array measurements of Kukula et al. (1995), which give a
comparable spatial resolution over the redshift range of the CfA sample
($0.002 \leq z \leq 0.070$) to our current A-array observations over
the range covered by the RQQs ($0.035 \leq z \leq 0.290$). Thus, in
both sets of data roughly the same amount of the host galaxy has been
included in the primary beam.  It is important to note, however, that,
unlike the CfA Seyfert sample, the current RQQ sample is in no sense
complete, nor are the selection criteria of the RQQs and Seyferts
comparable.

There is some overlap in terms of absolute magnitude between the two
samples. Assuming neutral $B-V$ colours for the quasars and Seyfert
nuclei, such that $M_{V} \simeq M_{B}$, it is clear that the two
lowest redshift RQQs in our sample, PG~0921+525 and PG~1149$-$110,
fall below the optical luminosity threshold for quasars and should,
strictly speaking, be regarded as Seyferts 1s. Equally, two of the
more distant CfA Seyferts (Mrk~205 \& Mrk~231) should strictly be
regarded as RQQs, since they have M$_{\rm V} < -23$. One object
(PG~0050+124; I~Zw~1) is common to both samples and is treated here as
an RQQ. For simplicity, in the discussion which follows we regard
everything with $M_{V} > -23$ as a Seyfert galaxy and everything with
$M_{V} < -23$ as a quasar.

The data in Figure~4 do appear to show a broad correlation between
radio luminosity and absolute magnitude. In the case of the CfA Seyfert
galaxies the correlation is already well known from the 5-GHz observations
of Edelson (1987). Interestingly, however, the new data for RQQs puts
them on a natural extension of this correlation.

Three objects stand out from the general distribution. These are the
RQQs PG~0007$+$106 and 1635$+$119, both of which have compact,
flat-spectrum radio nuclei, and the `Seyfert 1' Mrk~231, which is
known to be heavily reddened.  However, as argued in Section~5.7,
PG~0007+106 and 1635+119 are radio-intermediate quasars, with
properties quite distinct from those of the general RQQ
population. The available evidence suggests that they may be
relativistically beamed sources, in which case their intrinsic radio
luminosities would be considerably lower than the measured
values. Meanwhile Boksenberg et al. (1977) estimate 2 magnitudes of
extinction towards the optical nucleus of Mrk~231 and applying this
correction moves the object to the right in Figure~4, placing it back
on the radio-optical correlation. On these grounds, we exclude these
three objects from our discussion (although we note that their
inclusion would not alter our conclusions).

There is a slight tendency for the lower edge of the distribution to be
defined by radio upper limits rather than detections. This is expected
since both the Seyfert and RQQ samples are derived from surveys with
an optical flux limit and will therefore suffer to some extent from an
artificial correlation between $M_{V}$ and redshift. Nevertheless it
is worth noting that the optical and radio flux limits are different
for the two samples, offering us a degree of independence, and in any
case the number of upper limits is relatively small compared to the
total number of objects (9/46). This is compelling evidence in favour
of there being a real correlation between the radio and optical
luminosities even though the true distribution may not be as tightly
constrained as Figure~4 suggests at first glance.

We can derive a probability that the correlation is real, taking into
account all of the radio upper limits as well as the detections, by
employing the techniques of survival analysis (Feigelson \& Nelson
1985, Isobe, Feigelson \& Nelson 1986).

Using the {\sc asurv} package (Isobe \& Feigelson 1990), and
considering only the Seyfert~1s, we find the probability of a
correlation to be 98\% from Cox's proportional hazard model (supported
by the generalised Kendall's Tau and Spearman's Rho tests). This
probability increases to 99.99\% when the RQQs are included (there is
no significant correlation amongst the RQQs alone).

The parametric EM algorithm gives a regression line for the two
samples combined which is of the form 
\begin{equation}
\log(L_{8.4GHz}) = (-0.4\pm0.1)M_{V} + (12.9\pm1.3) 
\end{equation}
The fit is confirmed by the (non-parametric)
Buckley-James method and is shown as a dotted line in Figure~4.  When
the Seyfert sample is considered in isolation the resulting regression
line parameters are the same to within the calculated uncertainties.

Although the lower envelope of the radio luminosity distribution is
artificially imposed by the flux limits in the two sets of
observations, the upper envelope does seem to be real.  The correlation
appears to be telling us that for a given optical luminosity there is
a maximum radio luminosity which can be achieved by a radio-quiet AGN
(without the benefit of beaming effects) and that this maximum radio
luminosity {\it increases as a function of optical power}. This
implies a close link between the processes responsible for the radio
and optical emission.

As a caveat it should be noted that the measured optical luminosities
of the quasars and the Seyferts contain contributions from both the
active nucleus {\it and} the host galaxy. We cannot entirely rule out
the possibility that the radio emission is tied to the stellar
component of the host rather than the central engine, but there are
good reasons to suppose that the size of the stellar contribution is
not large enough to cause the observed correlation.  The $V$
magnitudes of the CfA Seyferts were measured over the nuclear region
of the galaxy so the fraction due to starlight, although ill-defined,
is not likely to be large (Huchra \& Burg 1992). For the quasars the
ratio of AGN to host luminosity for objects in this redshift and
magnitude range is typically 10:1 or greater in $B$-band (Taylor \&
Dunlop 1997), and so once again the contribution due to stars is
probably small.  A more accurate separation of host and quasar light
at optical wavelengths forms one of the goals of a forthcoming Cycle~6
Hubble Space Telescope program involving 12 of the RQQs in the current
sample.

For the moment it seems safe to assume that the observed distribution
reflects a correlation between the radio emission and the non-stellar
light from the quasar itself. Even so, recent studies have indicated
the existence of a correlation between quasar luminosity and host
galaxy mass ({\it eg} McLeod \& Rieke 1994), so the relationship
between radio emission and optical quasar emission in the present work
might simply reflect the fact that the most optically luminous quasars
occur in the most massive galaxies.

As a final point we note that it would be very interesting to see
whether the radio-optical correlation applies to more powerful RQQs
than those discussed here.  Our current sample consists of relatively
weak ($M_{V}>-26$) quasars, but if the correlation persists out to the
highest known absolute magnitudes for RQQs ($M_{V}\sim -30$) then
these objects would have typical radio luminosities of the order of
$L_{8.4GHz} \simeq 10^{24.5}$~W~Hz$^{-1}$. Kellermann et al.'s (1989)
radio survey of the BQS appears to support this, but the BQS is known
to be selectively incomplete for radio-quiet objects ({\it eg} Goldschmidt
et al. 1992) and so confirmation of this result will require
high-sensitivity radio surveys of carefully constructed quasar
samples.

\subsection{Comparison with Compact Steep-spectrum Sources}

The typical radio source dimensions of our RQQs ($\ll 10$~kpc) put
them in the same size regime as the Compact Steep-spectrum Sources
(CSSs) which make up between 15 and 30\% of the objects in radio
source catalogues (Fanti et al. 1990). There are, however, several
important differences between CSSs and the radio sources in RQQs. The
former, whose optical counterparts can be either elliptical galaxies or
quasars, are all radio {\it loud} (indeed, they are disproportionately
found towards the top end of the radio luminosity range), with radio
luminosities several orders of magnitude larger than anything in our
RQQ sample. The majority of quasar CSSs are also optically luminous
($M_{V} \ll -24$) and have high redshifts: there are no known objects
with $z < 0.2$.

One interpretation of CSSs is that they are {\it young} objects (Fanti
\& Spencer 1996), still in the process of `drilling' their way through
the interstellar medium of the host galaxy before escaping and
evolving into the giant radio sources typical of RLQs and RGs of
Fanaroff-Riley Type II (Fanti et al. 1990). In this model, as the
radio source propagates outwards from scales of a few hundred parsecs
to several kiloparsecs, changes in the radiation efficiency of the jet
will cause the total radio luminosity of the source to decrease by a
factor of $\sim 10$ (Fanti et al. 1995; Readhead et al 1995).

Are the compact RQQ radio sources in the current study also young
objects in the process of boring their way out of the surrounding
galaxy?  If so, do they ever break free, and what do they look like
when they do?  If we assume that the compact sources in RQQs undergo a
similar drop in radio luminosity as they expand from sub-kiloparsec to
galactic scales then the fully evolved, large-scale radio structures
would be too faint and diffuse to show up in our current maps.  Highly
sensitive, targeted observations of individual objects would be
required in order to detect such structures in RQQs.

Diffuse, bipolar radio structures have been observed in some Seyfert
galaxies on scales of tens of kiloparsecs - comparable to the size of
the host galaxy ({\it eg} Baum et al. 1993, Colbert et al. 1996). However,
compact, nuclear radio sources are also present in these objects so
the weak large-scale features are unlikely to be fully-evolved radio
sources, powered directly by the active nucleus in the manner of
radio-loud objects.  The structures are generally aligned with the
minor axis of the host galaxy, rather than that of the nuclear radio
source, and can be interpreted as plasma - perhaps a relic of previous
cycles of activity - which has lost its bulk kinetic energy and is now
drifting out of the plane of the host along the local density gradient
({\it eg} the features in Mrk~6; Kukula et al. 1996).

The fact that we detect compact radio components in 74\% of our sample
implies that there are few, if any, RQQs which have evolved into
large-scale radio sources.  This argues strongly that the radio
sources in RQQs remain confined to the central regions of the host
galaxy, at least over the lifetime of the optical quasar.

\subsection{The nature of the Radio-Intermediate Quasars}

Two of the objects in the current sample, PG~0007+106 (III~Zw~2) and
1635+119, possess unusually large radio luminosities and lie well
above the general distribution of radio-quiet quasars in the
radio-optical luminosity plane (Figure~4) (but still below that of
radio-loud quasars). MRS93 and Falcke, Sherwood \& Patnaik (1996)
describe such objects as `radio-intermediate quasars' (RIQs) and
interpret them as objects in which the radio emission is
relativistically boosted along the line of sight. The variable radio
and optical emission of these objects lends weight to such a scenario.

Several properties of the RIQs in our sample support this
hypothesis. The `genuine' RQQs invariably have steep radio spectra (at
least, whenever simultaneous flux measurements are available) with
typical spectral indices $\alpha^{1.4}_{8.4} \sim 0.7$. The two RIQs
stand out from this trend: 1635+119 has a flat radio spectrum, and
that of PG~0007+106 appears to be inverted.  Although we do not have
simultaneous flux measurements for PG~0007+106, its spectral shape is
confirmed by the observations of Schnopper et al. (1978) and
Gower \& Hutchings (1984).

A comparison of the current data with the observations of Gower \&
Hutchings (1984) shows that both sources are variable at radio
wavelengths. PG~0007+106 is also known to be highly variable in the
optical regime ({\it eg} Ter\"{a}sranta et al. 1992, Lainela 1994).

Brightness temperatures exceeding $10^{12}$K would constitute strong
evidence in favour of the RIQs being beamed sources, but the angular
resolution of the current VLA maps is much too low to place such a
stringent limit on $T_{B}$.  However, Lawrence et al. (1985) have
detected a VLBI component in PG~0007+106 less than 1~mas in size with
a measured brightness temperature at 22~GHz of $\sim 10^{11}$K.

PG~0007+106 and 1635+119 are 2 to 3 orders of magnitude more luminous
than the RQQs in our sample with absolute magnitudes $M_{V}\simeq -23$
(Figure~4).  Assuming that the jets in these objects are aligned very
closely to the line of sight, Lorentz factors of $\gamma \sim 5$ would
be sufficient to boost them from the general RQQ population to their
observed positions on the radio luminosity - absolute magnitude
plane. If this interpretation is correct then misaligned relativistic
jets should be a common feature of RQQs.

However, in both these objects the compact flat-spectrum core appears
to be associated with a radio lobe several arcseconds from the quasar
nucleus. The projected physical sizes of these two radio sources are,
at 33~kpc for PG~0007+106 and 175~kpc for 1635+119, much larger than
those of any of the RQQs (with the possible exceptions of PG~0804+761
and PG~1116+215). If the properties of the radio cores in PG~0007+106
and 1635+119 are indeed the result of beaming then (in the absence of
significant bending of the jets) this would imply that the radio
source axis is very close to the line of sight and that the actual
deprojected sizes are even larger.  Thus, even if the nuclear radio
components of the RIQs have unbeamed luminosities similar to those of
the genuine RQQs in the sample, the overall sizes of the radio sources
in these two objects are more in line with those observed in
radio-loud quasars.  In this case the RIQs might better be thought of
as RLQs of very low luminosity rather than RQQs which we happen to
observe from a preferred direction. Detailed investigations of a
larger sample of these `intermediate' objects will be required to
determine whether they ultimately bear more similarity to radio-loud
or radio-quiet quasars.

\subsection{FIR-radio correlation}

\begin{figure}
\vspace {10.0truecm}
 
\includegraphics{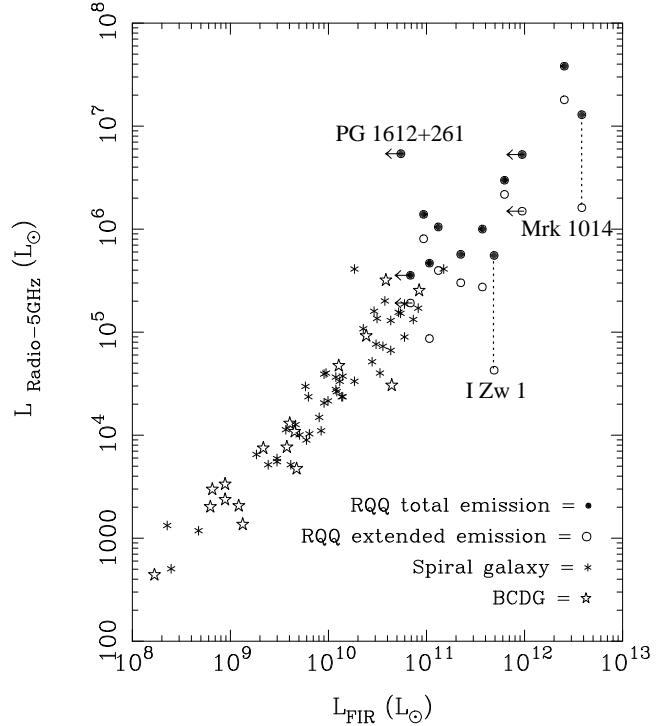}

\caption[]{5-GHz radio luminosity vs. 60-$\mu$m FIR (IRAS) luminosity
(both in units of solar luminosities) for RQQs in the current
sample. Filled circles represent the total ({\it ie} compact plus
extended) radio emission as measured by Kellermann et al. (1989) with
the VLA in D-configuration, whilst open circles represent the purely
extended emission (total emission minus the compact nuclear emission
from our A-array maps). IRAS upper limits are indicated by
left-pointing arrows.  Also shown are data for spiral galaxies
(asterisks) and Blue Compact Dwarf Galaxies (stars) from Wunderlich et
al. (1987).}

\end{figure}

If a significant fraction of the radio emission from RQQs can be
accounted for by a non-thermal source associated with the central
engine then we are left with the problem of why they appear to lie on
the far-infrared/radio correlation defined by galaxies which are
undergoing starformation.

In high-mass starforming regions young OB stars heat their surrounding
dusty, molecular material and consequently such regions emit strongly
at FIR wavelengths. These massive stars, which evolve off the main
sequence within $\sim 10^{7}$ years, are also the progenitors of the
supernova remnants which accelerate electrons and produce radio
synchrotron emission.  Hence starformation is understood to be the
origin of the tight correlation between the FIR and extended radio
luminosities in the disks of spirals and starburst galaxies (Helou,
Soifer \& Rowan-Robinson 1985) over at least four decades in
luminosity. In Figure~5 we show the 5-GHz radio luminosities of spiral
and blue compact dwarf galaxies plotted as a function of of their
60-$\mu$m IRAS luminosities (data from Wunderlich et al. 1987). In the
same figure we also show those RQQs from the current sample for which
IRAS fluxes were available.

The apparent overlap between the FIR and radio fluxes of RQQs and the
FIR-radio correlation for spiral galaxies and starbursts led Sopp \&
Alexander (1991) to suggest that both the FIR and radio luminosities
in RQQs were due entirely to starformation in the host galaxies and,
as Figure~5 demonstrates, most of the quasars in the current sample
(filled circles) do indeed appear to lie on this correlation. However,
in Figure~5 we have also plotted the position of RQQs on the FIR-radio
plane if only {\it extended} ({\it ie} galaxy-wide) emission is
considered (open circles). This helps to demonstrate a number of
important points which have perhaps not been properly appreciated
before.

First, the scatter in the FIR-radio relation is sufficiently large
that, in almost all cases, removal of the {\it dominant} contribution
of the compact nuclear radio source leaves the RQQ still in a location
consistent with the extension of the FIR-radio correlation of normal
star-forming galaxies. It is thus perfectly possible for the radio
emission from RQQs to be dominated by a non-thermal nuclear source,
and yet for them to still appear to lie on the FIR-radio correlation
displayed by normal and starburst galaxies. (One object, PG~1612+261,
lies conspicuously above the correlation and it is interesting to note
that in this instance {\it all} of the radio emission from the RQQ
appears to originate in the jet-like radio source in the nucleus.)

Second, two objects stand out on this diagram in that plotting only
their extended radio luminosities places them well below the FIR-radio
relation: PG~0050+124 (I~Zw~1) and PG~0157+001 (Mrk~1014). These two
sources can be used to illustrate how an RQQ can conspire to lie on
the FIR-radio relation even if its radio emission is almost totally
due to an active nucleus. If we suppose that their {\it extended}
radio luminosities are indicators of the level of starformation in the
host galaxies then their far-infrared luminosities can be viewed as an
order of magnitude larger than expected on the basis of of the
FIR-radio correlation.

This excess FIR luminosity must therefore be due to either (i)
additional {\it nuclear} starburst activity or (ii) additional dust
heating by the AGN. Our radio data offer a means to differentiate
between these two alternatives, because if the former option is
correct the compact radio source must also be due to the nuclear
starburst since its addition is just sufficient to return the source
to the FIR-radio correlation. However, as argued above, the spectral
indices and apparent brightness temperatures of the nuclear radio
sources in these two objects (as with many of the other quasars in our
sample), and (in the case of PG~0157+001) the structure of the radio
source, favour a non-starburst origin for the compact radio
components.

This in turn implies that the extra far-infrared luminosity of these
two sources is {\it not} due to a nuclear starburst but rather to
excess dust heating by the central engine. Interestingly, these are
the only two quasars in the current sample in which strong CO line
emission has also been detected.

If this interpretation is correct, then the fact that these sources
are lifted back onto the FIR-radio correlation by the addition of
their compact nuclear radio components must be due to a link between
the optical luminosity of the quasar nucleus (responsible for the dust
heating) and the radio emission from the AGN. This is entirely
consistent with the results of Section~5.5, in which we showed that
there does indeed seem to be a relation between the nuclear
optical and radio emission in RQQs and Seyfert~1 nuclei.

In conclusion the fact that most RQQs appear to lie in a natural
extrapolation of the the FIR-radio relation {\it cannot} be taken as
evidence that the radio emission is dominated by a starburst
component. Rather, the large scatter in this relation, coupled with
the fact that additional AGN heating of dust brings with it a
proportional amount of AGN-related radio emission means that {\it no}
RQQs are found to lie below the lower border of the relation.

\subsection{Summary of important results}

The angular resolution of the current observations is not sufficient to
unambiguously determine the nature of the nuclear radio emission in
RQQs. However, from the data we are able to place reasonably tight
constraints on the emission mechanisms and thus to make some strong
statements about the likely origins of the compact radio components.

To explain the radio luminosities of the compact components with a
pure starburst model would require supernova rates at least as high as
that of our own Galaxy (and in some cases much higher than the rates
observed in nearby starburst galaxies) occurring in a region no more
than a few hundred parsecs across. Lower limits on the brightness
temperatures of the deconvolved radio components are close to the
maximum values possible for starburst regions, but the radio spectra
are steep, indicating a predominantly non-thermal origin, contrary to
what one would expect if supernovae and H{\sc ii} regions were the
primary source of the emission.

The morphologies and sizes of the nuclear radio sources in RQQs bear
close comparison to those of Seyfert galaxies, and the RQQs fall on
the extension of the Seyfert 1 radio-optical luminosity correlation.
Imaging studies of nearby Seyferts have now resolved the nuclear
structure, in many cases revealing collimated radio jets, and it seems
overwhelmingly likely that such structures are also present in
RQQs. Forthcoming MERLIN and VLBI studies should confirm this.  The
high rate of occurrence of compact nuclear radio sources in RQQs
implies that the radio emission remains confined to the nucleus of the
host galaxy for a large proportion of the quasar's lifetime, {\it ie}
the radio sources are not small simply because they are young.

The properties of the two radio-intermediate quasars in the sample are
generally consistent with them being beamed versions of RQQs of the
same optical luminosity, suggesting that at least some RQQs also
possess relativistic jets.

Finally, we have shown that the tendency of RQQs to lie on the
FIR-radio correlation for star-forming spirals and starburst galaxies
cannot be taken as proof that stellar processes are responsible for
the radio emission in RQQs. Dust-heating by the quasar and associated
radio emission from the AGN can conspire to keep the RQQ on the
FIR-radio relation.

\section{Radio properties and the nature of the host galaxy in our host galaxy sample}

\begin{table*}\label{matched}
\caption{Objects in our host galaxy sample of RQQs. Here we list the
luminosities, structures and spectral indices of the radio sources in
these RQQs, along with the $K$-band absolute magnitude, scalelength
and type of their host galaxies as determined by Taylor et
al. (1996). Host morphology is defined as either predominantly
elliptical (E), disc-like (D) or ambiguous (?). The two RIQs are
denoted by asterisks.}

\centering
\begin{tabular}{rrrcccc}
\hline
 & \multicolumn{3}{c}{Radio source properties} & \multicolumn{3}{c}{Host galaxy properties} \\
Source & log(L$_{4.8}) $ & $\alpha^{1.4}_{8.4}$ & Structure & $M_{K}$ & Scale & Type\\ 
 & (W~Hz$^{-1}$) &  & & (2.2$\mu$m) & (kpc) &  \\ \hline
PG 0007+106 &$^{*}$24.69&$-$0.6&core-lobe&$-$24.47&15.4&  D\\
0046+112    & 23.47  & 0.7   & point & $-$23.74&     --&  ?\\
PG 0052+251 & 22.65  & 0.3   &core-jet?&$-$24.89&    18&  E\\
0054+144    & 22.94  & 0.6   & point & $-$26.13&     15&  E\\
PG 0157+001 & 23.82  & 0.7   & double& $-$26.66&     20&  E\\
0244+194    &$<$22.44&  --   & -     & $-$24.72&      9&  D\\
0257+024    & 23.52  & 0.5   & point & $-$25.56&     10&  D\\
PG 0923+201 &$<$22.72&  --   & -     & $-$25.25&     --&  ?\\
PG 0953+414 &$<$22.66&  --   & -     & $-$25.71&     --&  ?\\
PG 1012+008 & 23.05  & 0.9   & point & $-$26.13&   19.5&  D\\
PG 1211+143 &$<$22.38&  --   & -     & $-$24.43&    8.0&  D\\
PG 1440+356 & 22.31  &  --   & point & $-$25.16&    7.1&  D\\
1549+203    &$<$22.75&  --   & -     & $-$25.02&    46?&  D?\\
1635+119    &$^{*}$24.12&$-$0.02&core-lobe&$-$25.20& 28&  E\\
PG 2130+099 & 22.48  & 1.8   & triple& $-$25.28&    7.5&  E\\
2215-037    &$<$22.82&  --   & -     & $-$25.46&    13?&  D?\\
2344+184    &$<$22.19&  --   & -     & $-$25.11&     12&  E\\
\hline
\end{tabular}

\end{table*}

Taylor et al. (1996) have determined the NIR ($K$-band: 2.2$\mu$m)
luminosities, scalelengths and morphological types for the host
galaxies surrounding 17 of the RQQs in the current sample (Table~7).
They found that, whilst $\sim 50\%$ of the RQQs were found in disc
systems, for a significant fraction of objects the best fitting model
was in fact an {\it elliptical} galaxy. This is consistent with the
results of several other studies which also find RQQs to be
distributed amongst a mixture of disc-like and spheroidal hosts
(eg. V\'{e}ron-Cetty \& Woltjer 1990, Disney et al. 1995). However,
Taylor et al. also found a tendency for the quasars in disc hosts to
be less luminous than those in ellipticals. Although the sample was
too small for the result to be statistically significant, the trend is
supported by the results of McLeod \& Rieke (1994), who determined the
NIR morphologies of galaxies containing low-luminosity RQQs close to
the RQQ/Seyfert boundary and found that essentially all of these
objects lie in disc systems. Taylor et al. speculate that for $M_{V}
\leq -26$ all RQQs might occur in elliptical galaxies rather than
discs.

The AGN traditionally associated with elliptical hosts ({\it ie} radio
galaxies and RLQs) produce large, powerful radio sources and this
suggests that the RQQs in elliptical galaxies, whilst technically
`radio quiet', might still harbour radio sources which differ in size
and luminosity from those in disc galaxies.

Within our sample of 17 RQQs, 8 objects have hosts in which the
dominant stellar component is disc-like, 6 lie in spheroidal systems
and 3 have ambiguous morphologies.  We find {\it no} statistically
significant differences between the radio sources in different types
of host galaxy. The size distributions of the radio sources (including
upper limits) are indistinguishable for elliptical and disc hosts.
The RQQs in elliptical hosts {\it do not} contain more luminous radio
sources than those in disc galaxies, although this may not be true for
values of $M_{V} \ll -26$ if all optically luminous RRQs are found in
ellipticals {\it and} the radio-optical correlation is real.  Radio
spectral indices are generally steep in both types of host, and it
appears that radio-intermediate quasars can occur in either. If the
interpretation of these objects as beamed sources is correct then we
can infer that having an elliptical galaxy as a host is not a
prerequisite for the formation of a relativistic jet.

One interesting (though not formally significant) point is that, of
the 5 objects in Table~7 in which radio structure has been {\it
resolved}, 4 are in elliptical systems (the fourth is an RIQ and therefore
may not be representative of the radio-quiet population). This is the
only indication that RQQs in elliptical galaxies might have different
radio properties from those in discs.

The current sample is small, and is limited to RQQs of relatively low
optical luminosity, but our radio survey clearly demonstrates that
having an elliptical host does not automatically confer a large radio
luminosity on the quasar.  A significant number of elliptical galaxies
with active nuclei do {\it not} produce large, powerful radio
sources. Instead they contain small, weak sources very similar to
those in disc systems. More detailed studies of the host galaxies,
such as our own $R$-band HST imaging programme, should allow us to
determine if and how these elliptical RQQ hosts differ from the
elliptical systems containing radio-loud AGN.

\subsection{Radio source morphology and related structures 
in the host galaxy}

In two of the RQQs with resolved radio morphologies we also see
evidence for structure in the host galaxies along similar position
angles, and thus possibly related to the radio emission.

In PG~0157$+$001 the two radio `lobes' straddle the optical nucleus in
the same PA as an emission-line structure visible in the [O{\sc iii}]
image by Stockton \& MacKenty (1987). The scale of the [O{\sc iii}]
emission is $10''$ (37~kpc) - much larger than the $2''$ (7~kpc)
separation of the radio lobes, so the two features are unlikely to be
directly associated. However, the morphology is consistent with standard
models of anisotropy in AGN in which radio jets are ejected along the
axis of a cone of radiation from the central engine which can ionise
gas clouds further out in the body of the host galaxy.

The triple radio source in PG~2130+099 is perpendicular to the major
axis of the host (as determined by Taylor et al. 1996) but is aligned
with an anomalously blue region of the galaxy nucleus (Hutchings \&
Neff 1992). Once again, this is consistent with the standard model for
AGN, in which quasar light is emitted preferentially in the direction
of the radio jet but is scattered back into the line of sight by
material in the host galaxy. Alternatively, the blue excess may be due
to a burst of massive starformation triggered by the radio jet as it
passes through the host. At 8.4~GHz we find marginal evidence for
extended emission across the quasar nucleus in a direction orthogonal
to that of the triple source - {\it ie} parallel to the major axis of
the host galaxy (Figure~2). This is similar to radio structure in the
Seyfert nucleus of NGC~5929 (Su et al. 1996). Detection of a
high-frequency excess in PG~2130+099 at 15~GHz (Antonucci \& Barvainis
1988) lends support to the theory that the orthogonal structure might
consist of thermal emission from a circumnuclear starburst lying in
the plane of the host galaxy.  The emission would have to have a low
$T_{B}$ ($\leq 10^{5}$~K), but this should be testable with more
sensitive observations at 8.4 \& 15~GHz.

\section{Summary}

We have measured the radio fluxes and spectral indices of a sample of
27 RQQs using the VLA in A-configuration. The angular resolution of
the VLA is not sufficient to unambiguously determine the origin of the
radio emission in these objects, but from the data we are able to
conclude the following:

{\bf (1)} In 74\% of the sample we detect a radio source which is
coincident with the optical quasar to within the positional
uncertainty of the optical measurements.

{\bf (2)} The spectral indices of the RQQs are generally steep
($\alpha \sim 0.7$, where $S \propto \nu^{-\alpha}$), although two
objects, which also exhibit variability and unusually large radio
luminosities, have flat radio spectra.

{\bf (3)} Lower limits on the brightness temperatures of many of the
radio sources place them at the upper end of the range expected for
emission related to stellar processes ($T_{B} \sim 10^{5}$~K) ({\it
ie} in the regime in which we would expect to see some spectral
flattening if the emission was indeed stellar in origin). In some
objects the brightness temperature is almost certainly many orders of
magnitude greater than this, in which case the radio emission {\it
cannot} be produced by a starburst. Large numbers of supernovae
occurring within a very small volume would be required in order to
reproduce the observed radio luminosities.  Future observations with
greater angular resolution will place more rigorous constraints on all
these parameters.

{\bf (4)} In nine (possibly ten) objects we are able to resolve
radio structure, which takes the form of double, triple and linear
sources on scales of a few kiloparsecs. These structures are
consistent with the 100-pc-scale radio jets observed in
low-redshift Seyfert nuclei, and constitute strong evidence that
collimated ejection of radio plasma from the central engine is
occurring in RQQs.

{\bf (5)} The distribution of radio luminosities in RQQs form a
natural extension to those of Seyfert 1 nuclei. There appears to be a
correlation between radio luminosity and the optical absolute
magnitude of the quasar, suggesting a close relationship between the
central engine and the mechanism responsible for the bulk of the radio
emission.

{\bf (6)} There are no statistically significant differences in the
present data between the radio properties of the RQQs in
disc-dominated galaxies and those in elliptical hosts. 

Thus, our results are consistent with the conclusions of other recent
radio studies of RQQs: that in at least some RQQs a significant
contribution to the overall radio emission comes from a compact
nuclear source which is directly associated with the central engine
of the quasar, and which is qualitatively similar to the more powerful
radio sources observed in RLQs.

\section*{Acknowledgments}

The authors would like to thank Katherine Blundell for adding several
sources into her {\sc observe} file at very short notice, and Philip
Miller for allowing us to use three of the maps from his thesis. We
would also like to thank Ian Browne, the referee, for several very
useful comments and suggestions. MJK acknowledges PPARC support. This
research has made use of the NASA/IPAC Extragalactic Database (NED)
which is operated by the Jet Propulsion Laboratory, Caltech, under
contract with the National Aeronautics and Space Administration. The
National Radio Astronomy Observatory is operated by Associated
Universities, Inc., under cooperative agreement with the National
Science Foundation.

\end{document}